\documentclass[aps,prb,onecolumn,nofootinbib,citeautoscript,10pt]{revtex4-2}

\usepackage{amsmath,amssymb,bm} 
\usepackage{graphicx,comment}

\usepackage[tight]{subfigure} 

\usepackage[dvipsnames]{xcolor} 
\usepackage[papersize={8.5in,11in}]{geometry}
\usepackage[colorlinks=true]{hyperref}
\hypersetup{
    bookmarks=true,         
    unicode=false,          
    pdftoolbar=true,        
    pdfmenubar=true,        
    pdffitwindow=false,     
    pdfstartview={FitH},    
    pdfkeywords={keyword1} {key2} {key3}, 
    pdfnewwindow=true,      
    colorlinks=true,       
    linkcolor=magenta, 
    citecolor=blue,        
    filecolor=magenta,      
    urlcolor=blue           
} 

\usepackage{graphicx}
\usepackage{dcolumn}
\usepackage{color}
\usepackage{amssymb,amsmath}
\usepackage{tabularx,graphicx}
\usepackage{epstopdf}
\usepackage{latexsym}
\usepackage{colortbl}
\usepackage{psfrag}
\usepackage{bbm,bm,array,physics}
\usepackage{dsfont}
\usepackage{float, mathrsfs}

\def \nn{\nonumber \\}
\def\*#1{\mathbf{#1}} 

\begin{document}
\title{Linear response in planar Hall and thermal Hall setups for Rarita-Schwinger-Weyl semimetals}

\author{Rahul Ghosh}

\author{Firdous Haidar}

\author{Ipsita Mandal}
\email{ipsita.mandal@snu.edu.in}

\affiliation{Department of Physics, Shiv Nadar Institution of Eminence (SNIoE), Gautam Buddha Nagar, Uttar Pradesh 201314, India}

\begin{abstract} 
We investigate the nature of the linear-response tensors in planar Hall and planar thermal Hall setups, where we subject a Rarita-Schwinger-Weyl (RSW) semimetal to the combined influence of an electric field $\mathbf E $ (and/or temperature gradient $\nabla_{\mathbf r } T$) and a weak (i.e., nonquantizing) magnetic field $\mathbf B $. For computing the in-plane transport components, we have added an elastic deformation which gives rise to a chirality-dependent effective magnetic field $ \mathbf B^{\text{tot}} = 
\mathbf B  + \chi \, \mathbf B_5 $, where $\chi $ is the chirality of an RSW node. We have included the effects of orbital magnetic moment (OMM), in conjunction with the Berry curvature (BC), both of which appear as a consequence of nontrivial topology of the bandstructure. Due to the presence of four bands, RSW semimetals provide a richer structure for obtaining the linear-response coefficients, compared to the Weyl semimetals. In particular, we have found that the OMM-contributed terms may oppose or add up to the BC-only parts, depending on which band we are considering. Last, but not the least, we have determined the out-of-plane response comprising the intrinsic anomalous-Hall and the Lorentz-force-contributed currents, whose nature corroborates the findings of some recent experimental results.
\end{abstract}

\maketitle

\tableofcontents

\section{Introduction}

There has been an overwhelming amount of research activities involving the transport characteristics of semimetals, which are materials containing band-crossing points in the Brillouin zone (BZ) near the Fermi level, determined and protected by some symmetry. The terminology of ``semimetals'' originates from the fact that the density-of-states goes exactly to zero at these nodal points, showcasing a feature which is neither like insulators (as there is no gap) nor like conventional metals --- it lies somewhere in the middle of the two opposites. The simplest and the most well-known three-dimensional (3d) example is the Weyl semimetal (WSM) \cite{burkov11_Weyl, yan17_topological}, which harbours an isotropic linear-in-momentum dispersion in the vicinity of twofold band-crossings. The most straightforward generalization of the WSM is a semimetal with multifold band-crossings, where each band exhibits an isotropic linear dispersion. The low-energy effective Hamiltonian of a system, with $(2\, \varsigma + 1) $ bands touching at a point, can be expressed as $\sim \mathbf{k} \cdot \boldsymbol{\mathcal{S}} $, where $\boldsymbol{\mathcal{S}}$ represents the three components of the angular momentum operator in the spin-$\varsigma$ representation of the SO(3) group. This gives rise to emergent quasiparticles carrying pseudospin values equal to $ \varsigma  $. We use the nomenclature ``pseudospin'' in order to clearly distinguish this quantum number from the relativistic spin (arising from the spacetime Lorentz invariance).
Examples of multiband semimetals, featuring $\varsigma> 1/2$, include the pseudospin-3/2 Rarita-Schwinger-Weyl (RSW) semimetals~\cite{bernevig, long, igor, igor2, isobe-fu, tang2017_multiple, ips3by2, ips-cd1, ma2021_observation, ips-magnus, ips-jns, ips_jj_rsw, claudia-multifold} with fourfold band-crossings. 

In the realm of high-energy physics, the Rarita-Schwinger (RS) equation describes the field equation for elementary particles carrying the relativistic spin of 3/2, postulated in various supergravity models \cite{weinberg}. However, they neither appear in the standard model, nor have they been detected experimentally. On the other hand, in nonrelativistic condensed matter systems, we find that there are 230 space groups, paving the way for the emergence of a rich variety of unconventional excitations. In fact, for each one of these space groups, there exist pseudospin quantum numbers, dictated by the irreducible representations of the little group of lattice symmetries, at the high-symmetry points (in the BZ) \cite{bernevig}.
The RSW semimetals represent one such possibility, mimicking the relativistic spin-3/2 fermions, because of the quasiparticles carrying pseudospin-3/2. Their nomenclature thus mirrors the elusive high-energy RS fermions. 

To investigate the so-called \textit{topological} properties of a solid-state material, we consider the BZ as a closed manifold. When a nodal-point semimetal is said to possess a BZ endowed with a nontrivial topology, the nodes represent zero-dimensional topological defects, which thus carry nontrivial topological charges in the form of Berry curvature (BC) monopoles \cite{fuchs-review, polash-review}. The sign of the monopole charge gives us the chirality $\chi$ of the node, leading to the nomenclature of \textit{right-moving} and \textit{left-moving} \textit{chiral} quasiparticles, corresponding to $\chi = 1$ and $\chi = -1$, respectively. The net monopole charge, summed over all the nodes in the BZ, is constrained to vanish, which is easily explained by invoking the Nielson-Ninomiya theorem \cite{nielsen81_no}. We will adopt the convention that $\chi$ refers to the sign of the monopole charges (or, equivalently, the Chern numbers) of the negative-energy bands (i.e., the valence bands) --- thus a positive (negative) sign indicates that the node acts as a source (sink) for the flux lines of the BC vector field. For a given band, the monopole charge is obtained by integrating the BC flux over a closed two-dimensional (2d) surface enclosing the nodal point. The combined Chern number of all the bands corresponds to the wrapping of a generalized Bloch sphere $S^n$ (generalized to an $n$-level quantum system), representing the manifold of the quantum states. On the other hand, if we project on to the bands of a given pseudospin magnetic-quantum-number value (thus giving us a two-level system), the Chern number represents a wrapping number of the map from the 2d closed surface (topologically equivalent to $S^2$) to the Bloch sphere ($S^2$), given by the elements of the second homotopy group $\Pi_2(S^2) = \mathbb{Z}$. The WSM belongs to this category, with Chern numbers $\pm1$. This explains why the monopole charges, representing point defects, can be interpreted as Chern numbers as well.

Analogous to the WSMs, the RSW nodes carry nonzero values of the BC monopoles. The four bands at a single RSW node
have Chern numbers $\pm 1$ and $\pm 3$, which indicates a net monopole charge of magnitude 4. 
The indications of the existence of RSW quasiparticles have been linked to the large values of the topological charges detected in a range of materials, such as $\mathrm{CoSi}$ \cite{tang2017_multiple, takane2019_observation}, $\mathrm{RhSi}$ \cite{sanchez2019_topological},
$\mathrm{AlPt}$ \cite{schroter2019_chiral} and $\mathrm{PdBiSe}$ \cite{lv2019_observation}.
They also feature multiple Fermi arcs associated with the topological charge being an integer of magnitude $>1$.
The chiral anomaly, a signature property of the relativistic Weyl fermions, explained by Adler-Bell-Jackiw \cite{adler, bell}, continue to hold in the nonrelativistic settings involving WSMs \cite{chiral_ABJ}. In the context of the condensed matter systems, the anomaly refers to the phenomenon of charge pumping from one node (with chirality $\chi $) to its conjugate (with chirality $- \chi $), under the combined influence of applied electric ($\mathbf E$) and magnetic ($\mathbf B$) fields, with a rate $ \propto \mathbf E \cdot \mathbf  B $. Thus, for $\mathbf{E} \cdot \mathbf{B} \neq 0$, the number of quasiparticles of each chirality is not independently conserved in the vicinity of an individual node, and is known as the electrical chiral anomaly (ECA). Nevertheless, the net number of quasiparticles from the conjugate pairs of nodes in the entire BZ must yield zero, thus conserving the total electric charge. An analogous imbalance in chiral charge can be caused by adding (or replacing the external electric field by) an external temperature gradient $ \mathbf \nabla_{\mathbf r} T $, manifesting itself as the thermal chiral anomaly (TCA), being proportional to $ \nabla_{\mathbf r} T \cdot \mathbf  B $ \cite{das20_thermal}. A nonzero $ \nabla_{\mathbf r} T \cdot \mathbf{B} \neq 0$ also results in an imbalance in chiral energy (i.e., the energy carried by the quasiparticles of same chirality), which is sometimes referred to as the gravitational chiral anomaly (GCA) \cite{gca_andy, gca-expt, das20_thermal}, and it contributes to chiral energy-currents. 

\begin{figure}
\centering
\includegraphics[width=0.5 \linewidth]{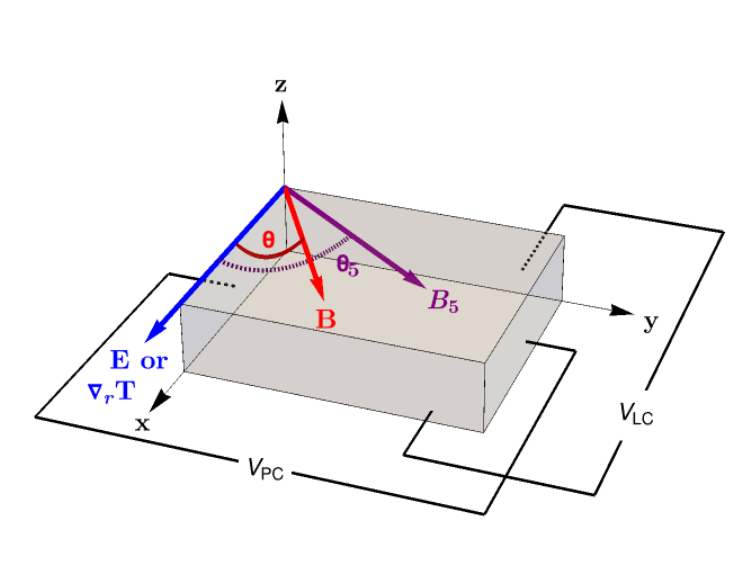}
\caption{Schematics of the planar Hall (or planar thermal Hall) setup, where an RSW node is subjected to a static electric field $ E\, {\mathbf{\hat x}} $ (or temperature gradient $\partial_x T\, {\mathbf{\hat x}}$). An external magnetic field $\mathbf B $ is applied at an angle $\theta $ with respect to the electric field (or temperature gradient). Additionally, the semimetallic slab is subjected to a mechanical strain, whose effect is captured via an artificial pseudomagnetic field $\mathbf B_5 $, making an angle $\theta_5$ with the $x$-axis. The in-plane voltage generated parallel and perpendicular to $ E\, {\mathbf{\hat x}} $ (or $\partial_x T\, {\mathbf{\hat x}}$) are indicated by $V_{\rm LC}$ and $V_{\rm PC}$, respectively. The subscripts indicate their association with the longitudinal and Hall components of the resulting currents.
\label{figsetup}}
\end{figure}

In this paper, we consider planar Hall setups consisting of $\mathbf E $ (and/or $\nabla_{\mathbf r} T$) and $\mathbf B $ fields [cf. Fig.~\ref{figsetup}], and involving RSW semimetals. We choose $\mathbf E$ and $\mathbf B $ to lie in the $xy$-plane, with $ \mathbf E $ (or $\nabla_{\mathbf r} T$) applied along the $x$-axis. In other words, $\mathbf B =  B \left( \cos \theta, \sin \theta, 0 \right) $ (where $B \equiv |\mathbf B|$), such that $\theta \neq \pi/ 2 \text{ or } 3\,\pi/2$ in general.
In the linear response regime (with respect to $\mathbf E $ and $\nabla_{\mathbf r} T$), the node-dependent transport coefficients, relating the electric current to $\mathbf E $ and $\nabla_{\mathbf r} T$, are the magnetoelectric conductivity tensor ($\sigma^\chi$) and the magnetothermoelectric conductivity tensor ($\alpha^\chi $), respectively. A third response tensor, which we denote by $\ell^\chi $, is the linear response tensor relating the heat current to the temperature gradient at a vanishing electric field. Since $\ell^\chi $ contributes to the magnetothermal conductivity tensor $\kappa^\chi $, we will often loosely refer to $\ell^\chi $ itself as the magnetothermal coefficient. The longitudinal components, $\sigma_{xx}^\chi$ and $\alpha^\chi_{xx}$, are known as the longitudinal magnetoconductivity (LMC) and the longitudinal thermoelectric coefficient (LTEC), respectively. The transverse components, $\sigma_{yx}^\chi$ and $\alpha^\chi_{yx}$, are referred to as the planar Hall conductivity (PHC) and the transverse thermoelectric coefficient (TTEC), respectively. In recent times, there has been a surge in theoretical and experimental efforts to investigate various aspects of these response tensors~\cite{zhang16_linear, chen16_thermoelectric, nandy_2017_chiral, nandy18_Berry, amit_magneto, das20_thermal, das22_nonlinear, pal22a_berry, pal22b_berry, fu22_thermoelectric, araki20_magnetic, mizuta14_contribution, ips-serena, timm, onofre, ips_rahul_ph_strain, rahul-jpcm, ips-kush-review, claudia-multifold, ips-ruiz, ips-tilted}. 

All the linear-response coefficients invariably contain the information about nontrivial band topology via the inclusion of the BC. In addition, the orbital magnetic moment (OMM) \cite{xiao_review, sundaram99_wavepacket}, which is another artifact of a nontrivial topology in the bandstructures, also affects the behaviour of the response tensors \cite{timm, onofre, ips-ruiz}. Hence, in this paper, we include the effects of both the BC and the OMM, which constitute a complete description conveying the role of topology. It is worth mentioning here that complementary evidence of nontrivial topology in bandstructures, extensively explored in the literature, include intrinsic anomalous-Hall effect~\cite{haldane04_berry,goswami13_axionic, burkov14_anomolous}, magneto-optical conductivity when Landau levels are relevant~\cite{gusynin06_magneto, staalhammar20_magneto, yadav23_magneto}, Magnus Hall effect~\cite{papaj_magnus, amit-magnus, ips-magnus}, circular dichroism \cite{ips-cd1, ips_cd}, circular photogalvanic effect \cite{moore18_optical, guo23_light,kozii, ips_cpge}, and transmission of quasiparticles across potential barriers/wells \cite{ips_aritra, ips-sandip, ips-sandip-sajid, ips-jns}.

In addition to the action of the externally-applied magnetic field, we consider the case when the semimetal is subjected to a mechanical strain, thus undergoing elastic deformations. The effects of these deformations on the chiral quasiparticles can be modelled as pseudogauge fields \cite{guinea10_energy, guinea10_generating,low10_strain, landsteiner_gaguge,liu_gauge, pikulin_gauge, arjona18_rotational, onofre}, with the matter-gauge-field coupling constant being proportional to $\chi $ \cite{landsteiner_gaguge, liu_gauge,pikulin_gauge, ghosh20_chirality,girish2023, onofre}. Due to the chiral nature of the coupling between the emergent vector fields and the itinerant fermionic particles, it provides a platform to study the interactions of matter with axial vector fields in three dimensions. This property differs from the case of the actual electromagnetic fields, where the coupling constant is independent of $\chi$. It has been shown theoretically that a uniform pseudomagnetic field $ \mathbf B_5$ can be generated when a semimetallic nanowire is put under torsion \cite{pikulin_gauge}. Some direct experimental evidence for the generation of such pseudomagnetic fields is also available \cite{exp_gauge}.

The paper is organized as follows. In Sec.~\ref{secmodel}, we outline the form of the low-energy effective Hamiltonian in the vicinity of an RSW node, and the resulting expressions for the BC and the OMM. Secs.~\ref{secsigma} and \ref{secalpha} are devoted to demonstrating the explicit expressions for the longitudinal and transverse components of $\sigma^\chi $, $\alpha^\chi$, and $\ell^\chi$, respectively.
In Sec.~\ref{secah_lf}, we discuss the effects of the so-called Lorentz-force parts, which show up in the out-of-plane components of the response coefficients, and compare it with some recent experimental observations. Finally, we conclude with a summary and outlook in Sec.~\ref{sec_summary}. The appendices are devoted to elaborating on much of the details of the intermediate steps, necessary to derive the final expressions shown in the main text. 
In all our expressions, we resort to using the natural units, which implies that the reduced Planck's constant ($\hbar $), the speed of light ($c$), and the Boltzmann constant ($k_B $) are set to unity.

\begin{figure*}[t]
\subfigure[]{\includegraphics[width=0.35 \linewidth]{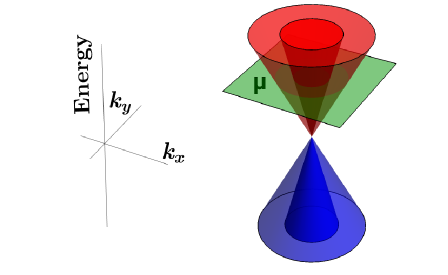}} \hspace{ 2 cm}
\subfigure[]{\includegraphics[width=0.2 \linewidth]{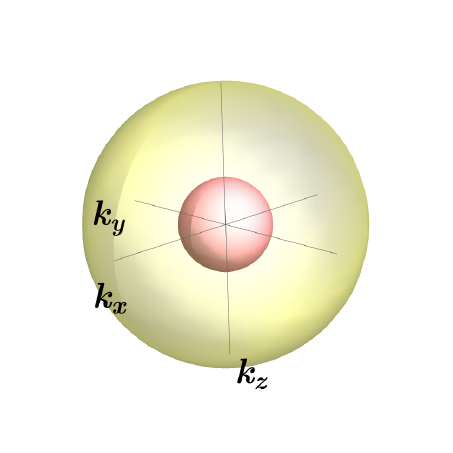} 
\includegraphics[width=0.2 \linewidth]{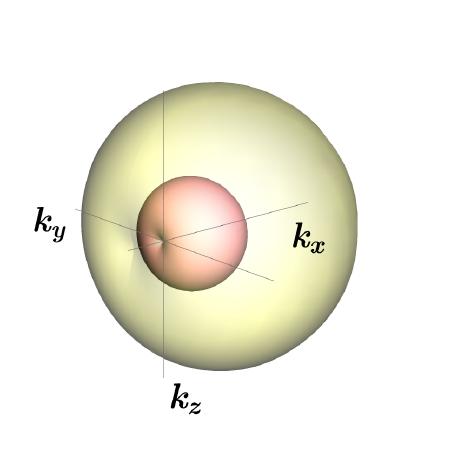}}
\caption{\label{figdis}(a) The dispersion of a single RSW node against the $k_x k_y$-plane. The chemical potential $\mu >0 $ cuts the conduction bands.
(b) Schematics of the Fermi surfaces at the node with chirality $+1$, without and with the OMM-correction for the effective energy dispersion. Here, we have taken the net effective magnetic field to be directed purely along the $x$-axis.}
\end{figure*}

\section{Model} 
\label{secmodel}

With the help of group-theoretic symmetry analysis and first principles calculations, it has been shown that seven space groups may host fourfold band-crossings \cite{bernevig} at high-symmetry points of the BZ. Nearly 40 candidate materials have also been identified that can host the resulting RSW quasiparticles.
Ref.~\cite{grushin-multifold} has tabulated the multifold degeneracies in the 65 chiral space groups characterizing the chiral crystals, which are the ones with only orientation-preserving symmetries. A chiral fourfold band-crossing can be realized in the space groups (1) 195–198 and 207–214 at the $\Gamma $-point; (2) 207 and 208 at the $R$-point; and (3) 211 and 214 at the $H$-point, in the presence of spin-orbit coupling. These fourfold degeneracies exhibit a BC texture that is homotopic to that of a spin-$3/2$ moment in a magnetic field. The RSW semimetal, with the effective Hamiltonian possessing a full SU(2) invariance (i.e., a full rotational invariance). It has been shown \cite{prb108035428} that chiral topological semimetals belonging to the SrGePt family (e.g., SrSiPd, BaSiPd, CaSiPt, SrSiPt, BaSiPt, and BaGePt), characterized by the space group 198, host RSW quasiparticles, sixfold excitations (two copies of pseudospin-1 fermions), as well as Weyl points in their bandstructures, when spin-orbit coupling is considered. More explicitly, a fourfold-degenerate node appears at the center of the BZ (i.e., the $\Gamma $-point), carrying the monopole charge of $+ \, 4 $, while a sixfold-degenerate node arises at the boundary of the BZ (i.e., the $R$-point) with a net monopole charge equalling $- \,2-2 = - \, 4$.

The usual method of linearizing the $\mathbf{k} \cdot \mathbf {p}$ Hamiltonian about such a fourfold-degeneracy point provides us with the low-energy effective continuum Hamiltonian, valid in the vicinity of the node. The explicit form of this Hamiltonian, for a single node with chirality $\chi $, is given by 
\begin{align}
	\mathcal{H}(\mathbf{k}) =  v_0\left( 
k_x\,	{\mathcal J }_x + k_y\, {\mathcal J }_y
+ \chi \, k_z \, {\mathcal J }_z \right),
\end{align}
where $ \boldsymbol{\mathcal J } = \lbrace {\mathcal J }_x,\, {\mathcal J }_y,\, {\mathcal J }_z \rbrace $ represents the vector operator whose three components comprise the the angular momentum operators in the spin-$3/2$ representation of the SU(2) group. We choose the commonly-used representation where
\begin{align}
{\mathcal J }_x= 
\begin{pmatrix}
	0 & \frac{\sqrt{3}}{2} & 0 & 0 \\
	\frac{\sqrt{3}}{2} & 0 & 1 & 0 \\
	0 & 1 & 0 & \frac{\sqrt{3}}{2} \\
	0 & 0 & \frac{\sqrt{3}}{2} & 0 
\end{pmatrix} , \quad
{\mathcal J }_y=
\begin{pmatrix}
	0 & \frac{-i \,  \sqrt{3}}{2}  & 0 & 0 \\
	\frac{i \, \sqrt{3}}{2} & 0 & -i & 0 \\
	0 & i & 0 & \frac{-i \, \sqrt{3}}{2}  \\
	0 & 0 & \frac{i \, \sqrt{3}}{2} & 0 
\end{pmatrix}, \quad
{\mathcal J }_z =
\begin{pmatrix}
	\frac{3}{2} & 0 & 0 & 0 \\
	0 & \frac{1}{2} & 0 & 0 \\
	0 & 0 & -\frac{1}{2} & 0 \\
	0 & 0 & 0 & -\frac{3}{2} 
\end{pmatrix}.
\end{align}

The energy eigenvalues are found to be
\begin{align}
\label{eqeval}
\varepsilon_s ( k ) =  s \, v_0 \, k \,, \quad 
s \in  \left \lbrace  \pm \frac{1}{2}, \pm \frac{3}{2} \right \rbrace,
\end{align}
where $ k = \sqrt{k_x^2 + k_y^2 + k_z^2 } $.
Hence, each of the four bands has an isotropic linear-in-momentum dispersion [cf. Fig.~\ref{figdis}(a)]. The signs of ``$+$'' and ``$-$'' give us the dispersion relations for the conduction and valence bands, respectively. The corresponding group velocities of the quasiparticles are given by 
\begin{align} 
\boldsymbol{v}_{s}(\mathbf{k}) &= 
\nabla_{\mathbf{k}}  \varepsilon_{s}(\mathbf{k})  
=  \frac{ s \, v_0 \,\mathbf k }{k} \, .  
\end{align}

\subsection{Topological quantities}

Due to a nontrivial topology of the bandstructure, we need to compute the BC and the OMM, using the starting expressions of \cite{xiao_review}
\begin{align} 
	{\mathbf \Omega}^{\chi}_{s}( \mathbf k) &=  i \, \langle \nabla_{ \mathbf{k}} \psi^{\chi}_{s} ({ \mathbf{k}})| \, \cross \, | \nabla_{ \mathbf{k}} \psi^{\chi}_{s} ({ \mathbf{k }}) \rangle 
\text{ and } 
	\mathbf{m}^{\chi}_{s}( \mathbf k) =
- \,\frac{e}{2} \, \text{Im} \left[ \langle \nabla_\mathbf{k} 
\psi^{\chi}_{s} \vert  \cross \left( \mathcal{H} {(\mathbf{k})} 
 - \varepsilon_{s}(\mathbf{k}) \right) 
	\vert  \nabla_\mathbf{k} \psi^{\chi}_{s} \rangle \right],
\end{align}  
respectively. Here, $| \psi^{\chi}_{s} ({ \mathbf{k }}) \rangle $ denotes the eigenfunction for the $s^{\rm th}$ band at the node with chirality $\chi $, and $e$ denotes the magnitude of the charge of a single electron. 
Evaluating these expressions for the RSW node described by $\mathcal{H}(\mathbf{k}) $, we get \cite{graf_thesis}
\begin{align}  
{\mathbf \Omega}^{\chi}_{s}( \mathbf k) &=    
	 - \, \frac{\chi \,   s \,\mathbf k  } {k^3}  \text{ and } 
\mathbf{m}^{\chi}_{s}( \mathbf k) 
= - \frac{e \, \chi\, v_0 \, \mathcal{G}_s  \,\mathbf k } {k^2} \, ,
\text{ where } \lbrace \mathcal{G}_{\pm 1/2}, \,\mathcal{G}_{\pm 3/2} \rbrace
 = \left \lbrace \frac{7}{4}, \, \frac{3}{4} \right \rbrace.
\end{align}
Since $\mathbf  \Omega^{\chi}_{s}( \mathbf k)$ and $\mathbf{m}^{\chi}_{s}( \mathbf k)$ are the intrinsic properties of the bandstructure, they depend only on the wavefunctions. Clearly, they are related as
\begin{align}
	\mathbf{m}^{\chi}_{s}( \mathbf k) = \frac{e \, v_0 \, \mathcal{G}_s \, k } {s} 
	\, {\mathbf \Omega}^{\chi}_{s}( \mathbf k) \,.
\end{align}
We observe that, unlike the BC, the OMM does not depend on the sign of the energy dispersion. 

The coupling between the OMM with the magnetic field gives rise to a Zeeman-like correction to the dispersion,
quantified by 
\begin{align} 
\label{eqomme}
\eta^{\chi}_{s} ({\mathbf k})  &= - \, \mathbf{m}^{\chi}_{s}  (\mathbf k) \cdot \mathbf{B}
=  e \, \chi\, v_0 \, \mathcal{G}_s  \,   
\frac{  \mathbf{k}   \cdot  \mathbf{B} } {k^2} \, .
\end{align}
Therefore, we have 
\begin{align} 
\label{eqtoten}
\xi^{\chi}_s ({\mathbf k}) & = \varepsilon_{s}(\mathbf{k})  
+ \eta^{\chi}_{s}(\mathbf{k}) 
\,, \quad 
\boldsymbol{w}^{\chi}_{s}(\mathbf{k})= 
	\boldsymbol{v}_s(\mathbf{k})  + \boldsymbol{u}^{\chi}_{s}(\mathbf{k}) \,,
\nn \boldsymbol{u}^{\chi}_{s}(\mathbf{k}) & =  \nabla_{\mathbf{k}} \eta^{\chi}_{s}(\mathbf{k})   
= e \, \chi \,v_0 \,  \mathcal{G}_s  \, \frac{ \mathbf{B}  
	-  2\, \hat{ {\mathbf k}} \left (\hat{  {\mathbf k}  } \cdot \mathbf{B} \right) }{k^2}	\,,
\end{align}
where $\xi^{\chi}_s ({\mathbf k})$ and $\boldsymbol{w}^{\chi}_{s}(\mathbf{k})$ are the OMM-modified energy and band velocity of the quasiparticles, respectively. With the usual usage of notations, $\hat{ \mathbf{k} }$ is the unit vector along $ \mathbf{k} $. The full rotational isotropy of the Fermi surface, for each band of the RSW node, is broken by the inclusion of the OMM corrections. This is depicted schematically in Fig.~\ref{figdis}(b) for the case when $\mathbf B$ is applied along the $ x $-axis.

\begin{table}[t]
\centering
\begin{tabular}{|c|c|c|}
\hline
Parameter &   SI Units &   Natural Units  \\ \hline
$v_0$ from Ref.~\cite{Nag_floquet_2020} & $15 \times10^{5} $ m~s$^{-1} $ & $0.005$  
\\ \hline
$\tau$ from Ref.~\cite{watzman18_dirac} & $ 10^{-13} \, \text{s} $ & $152  $ eV$^{-1}$  
\\ \hline
$ T $ from Refs.~\cite{claudia-multifold} & $ 10 - 100 $ K & $ 8.617 \cross 10^{-4} - 8.617 \cross 10^{-3}  $ eV
\\ \hline
$ B $ from Ref.~\cite{claudia-multifold} & $ 0 $ --- $ 10 \, \text{Tesla} $ 
& $ 0 $ --- $ 1950 \, \text{eV$^{2}$} $ \\ \hline
$\mu$ from Ref.~\cite{tang2017_multiple} & $  1.6\times 10^{-21} -  1.6\times 10^{- 20 } $ J 
& $0.01 $ --- $0.1$ eV \\ \hline
\end{tabular}
\caption{\label{table_params}The ranges of values for the various parameters, used in the plots of the linear-response coefficients, are tabulated here. While using the natural units, we need to set $\hbar=c=k_{B}=1$.}
\end{table}

\subsection{Linear-response coefficients}
\label{seclinear}

Let us investigate the transport properties in a planar Hall setup with an external magnetic field applied in the $xy$-plane, such that $\mathbf B = B \left( \cos \theta \,{\mathbf{\hat x}} + \sin \theta \, {\boldsymbol{\hat y}}\right) $. An electric field $\mathbf E = E \,  {\boldsymbol{\hat x}} $ and/or a temperature gradient $\nabla_{\mathbf r}  T = \partial_x T\, {\boldsymbol{\hat x}} $ are/is applied in a configuration co-planar with $\mathbf B$.
In the following two sections, we will compute the resulting three linear-response coefficients, $\sigma^\chi$, $\alpha^\chi$, and $\ell^\chi$, whose technical definitions can be found in Eq.~\eqref{eqcur1}. We will consider a positive chemical potential $ \mu $ being applied to the node, such that the Fermi level cuts the two conduction bands (labelled by $\tilde s$), which take part in transport. Consequently, we have here
\begin{align}
\sigma ^\chi = \sum \limits_{ \tilde s } 
\sigma^\chi_{\tilde s}\,, \quad
\alpha ^\chi = \sum \limits_{ \tilde s } 
\alpha^\chi_{\tilde s}\,,\quad
\ell^\chi = \sum \limits_{ \tilde s } 
\ell^\chi_{\tilde s}\,,
\text{ with } \tilde s \in \left \lbrace +\frac{1}{2}, \,+\frac{3}{2} \right \rbrace.
\end{align}

The steps to obtain the forms of linear-response coefficients have been reviewed in Appendix~\ref{secboltz}. Here, we assume that only the intranode scatterings are relevant, with a relaxation time $\tau$, and ignore any internode/interband scattering processes. The neglect of internode scatterings is justified if the energy offset between the fourfold-degenerate point (at $\Gamma$) and the sixfold-degenerate point (at $R$) is large \cite{tang2017_multiple, prb108035428, prl119206401, yamakage}.
From the solutions obtained in Appendix~\ref{appinter}, and setting $g_{\tilde s}= 1$ (i.e., ignoring the degeneracy due to electron's spin), we arrive at the following expressions for a single band of chirality $\chi$ and index $ \tilde s$:
\begin{align}
\label{eqsigmatot}
& \sigma_{\tilde s}^\chi 
= \sigma_{\tilde s}^{\chi, \rm AH}
+ \sigma_{\tilde s}^{\chi, \rm LF}
+ \bar \sigma_{\tilde s}^\chi \,, \quad
\left( \sigma_{\tilde s}^{\chi,\rm AH} \right)_{ij}
= -\, e^2  \,\epsilon_{ijl} 
\int \frac{ d^3 \mathbf k}{(2\, \pi)^3 } \, \left(\Omega^{\chi}_{\tilde s} \right)^l 
\,  f_0  (\xi^{\chi}_{\tilde s}) \,,\nn
& \left (\sigma_{\tilde s}^{\chi, \rm LF} \right)_{i j} = - \,\epsilon_{j q r} \, e^3 \, 
\tau ^2 \, {\tilde s}^3 \, v_0^3 
\int \frac{ d^3 \mathbf{k}} {(2 \, \pi)^3} 
\, \frac{({\mathcal{D}^{\chi}_{\tilde s}} )^2}
{ \left ( \varepsilon_{\tilde s} \right )^5}
 \left[ (w^{\chi}_{\tilde s})_i + (W^{\chi}_{\tilde s})_i \right ]  
  \left[ \left ( \varepsilon_{\tilde s} \right )^2 - \lambda_{\tilde s}^{\chi}\right ]^2  \,B_r \,
   \varrho_q \, f^\prime_0 (\xi^{\chi}_{\tilde s}) \,,\nn
&
\left(\bar \sigma^{\chi}_{\tilde s} \right)_{i j} 
= - \,e^2 \, \tau  
\int \frac{ d^3 \mathbf k}{(2\, \pi)^3 } \, \mathcal{D}^{\chi}_{\tilde s} 
\left[  (w^{\chi}_{\tilde s})_i \, + (W^{\chi}_{\tilde s})_i \right ]
\left [ (w^{\chi}_{\tilde s})_j \, + (W^{\chi}_{\tilde s})_j \right] \, f^\prime_0 (\xi^{\chi}_{\tilde s})  \,.
\end{align}
Here,
\begin{align}
\label{eqdist}
	f_0 \big (\xi^\chi_{\tilde s}(\mathbf k) , \mu, T (\mathbf r) \big )
= \frac{1}
{ 1 + \exp [ \frac{ \xi^\chi_{\tilde s}(\mathbf k)-\mu } 
			{ T  (\mathbf r )}  ]}
\end{align}
is the equilibrium Fermi-Dirac distribution of the quasiparticles at temperature $T$,
\begin{align}
 \mathbf{W}^{\chi}_{\tilde s} = 
  e \left  ( \boldsymbol{w}^{\chi}_{\tilde s} \cdot 
  \boldsymbol{\Omega}^{\chi}_{\tilde s} \right  ) \mathbf{B} \,,
 \end{align}
and
\begin{align}
 \boldsymbol{\varrho} =  \cos{\phi} \sin{\gamma}  \, {\boldsymbol{\hat x}}
 + \sin{\phi} \sin{\gamma}  \, {\boldsymbol{\hat y}} + \cos{\gamma}  \, {\boldsymbol{\hat z}} \,,
  \quad \lambda_{\tilde s}^{\chi} = 
  \vartheta \sum_i \varrho_i {B}_i \,,\quad
 \vartheta = 2 \, \chi \, e\, {\tilde s} \,  {\mathcal{G}}_{\tilde s} \, v_0^2 \,,
\end{align}
The variables $\phi $ and $\gamma $ refer to the azimuthal and polar angles of the spherical polar coordinates, which the components of $\mathbf k$ are transformed to, as shown in Appendix~\ref{appint}.
For the uncluttering of notations, we have suppressed the $\mu $- and $T$-dependence of $f_0$.
The ``prime'' superscript denotes differentiation with respect to the variable shown within the brackets [for example, $ f_0^\prime (u) \equiv \partial_u f_0 (u)$].
The three parts of $\sigma_{\tilde s}^\chi$ represent the following:
\begin{enumerate}

\item $\sigma_{\tilde s}^{\chi,\rm AH}$ gives the ``intrinsic anomalous-Hall effect'' \cite{haldane04_berry, goswami13_axionic, burkov14_anomolous}, with its longitudinal component being zero. This part is completely independent of the relaxation time $\tau $. If OMM is set to zero, $\sigma_{\tilde s}^{\chi,\rm AH}$ is independent of $\mathbf B$, and $  \sigma^{\chi,\rm AH}$
vanishes identically. We also note that, for our configuration with the applied fields and temperature gradient confined to
the $xy$-plane, the in-plane components (i.e., $ {xx} $- and $ {yx} $-components) are zero, and only the transverse out-of-plane component with $zx$-indices is nonzero. 

\item 
The second part is the so-called Lorentz-force part, and it arises from the current density
\begin{align}
{\mathbf J}_{\tilde s}^{\chi, \rm LF} =
 - \, e^3 \, \tau^2 \, {\tilde s}^3 \,v_0^3
  \int \frac{d^3 \mathbf{k} } {(2 \, \pi)^3} \, 
 \frac{({\mathcal{D}^{\chi}_{\tilde s}} )^2}
 {  \varepsilon_{\tilde s} ^5}  
 \left ( \boldsymbol{w}^{\chi}_{\tilde s} + \mathbf{W}^{\chi}_{\tilde s} \right )   
 \left [ \varepsilon_{\tilde s} ^2 
 - \lambda_{\tilde s}^{\chi} 
 \right ]^2 \, f^\prime_0 (\xi^{\chi}_{\tilde s})   
\left  (\boldsymbol{\varrho} \cross \mathbf{B} \right )  \cdot \mathbf{E} \,  .
\end{align}
The name has been coined to reflect the fact that it includes the classical Hall effect due to the Lorentz force.
The derivation for this part is quite tedious, but it has been detailed in Appendix~\ref{applor}.
The resulting $  {\sigma}^{\chi, \rm LF}_{\tilde s} $ contains only odd powers of $ B $.
Analogous to $\sigma_{\tilde s}^{\chi,\rm AH}$, its in-plane components are zero, and only the Hall component with $zx$-indices is nonzero.

\item The third part arises from the current density
\begin{align}
\bar {\mathbf J}^{\chi}_{\tilde s} &= 
- \, e^2 \, \tau 
\int \frac{ d^3 \mathbf{k}} {(2 \, \pi)^3} \, 
\mathcal{D}^{\chi}_{\tilde s}  \left[\boldsymbol{w}^{\chi}_{\tilde s} 
+ \boldsymbol{W}^{\chi}_{\tilde s} \right ]
\left[  \boldsymbol{w}^{\chi}_{\tilde s} \,+ \, \boldsymbol{W}^{\chi}_{\tilde s} \right ]  \cdot \mathbf{E} \;
 f^\prime_0 (\xi^{\chi}_{\tilde s}) \,,
\end{align}
and gives rise to $ \bar {\sigma}^{\chi}_{\tilde s}$, which contains only even powers of $ B $.

\end{enumerate}

For the magnetothermoelectric conductivity, we only consider the in-plane thermoelectric current density given by
\begin{align} 
\label{eqalphacur}
	\mathbf{ \bar J}^{\chi}_{\tilde s} &= 
	e \, \tau  \int \frac{d^3 \mathbf{k}} {(2 \, \pi)^3}
	  \, \mathcal{D}^{\chi}_{\tilde s} \left  (\xi^{\chi}_{\tilde s} - \mu_{\chi} \right )  
	\left (  \boldsymbol{w}^{\chi}_{\tilde s} \,+ \, \boldsymbol{W}^{\chi}_{\tilde s} \right ) 
\left[    \left (  \boldsymbol{w}^{\chi}_{\tilde s} +  \boldsymbol{W}^{\chi}_{\tilde s} \right )  
\cdot \frac{(-\nabla_{\mathbf{r}} T )}{T} \right ] 
	  f^\prime_0 (\xi^{\chi}_{\tilde s})\, ,
\end{align}
which gives rise to nonzero in-plane components. This
leads to the tensor components of
\begin{align} 
\label{eqalpha}
	\left (\bar{\alpha}^{\chi}_{\tilde s} \right )_{ij} &= 
e \, \tau  \int \frac{d^3 \mathbf{k}} {(2 \, \pi)^3} \,\mathcal{D}^{\chi}_{\tilde s}  
\left [ (w^{\chi}_{\tilde s})_i +  (W^{\chi}_{\tilde s})_i \right ]
\left [ (w^{\chi}_{\tilde s})_j +  (W^{\chi}_{\tilde s})_j \right ]
\frac{(\xi^{\chi}_{\tilde s} - \mu_{\chi} )}{T}
	   \, f^\prime_0 (\xi^{\chi}_{\tilde s}) \,.
\end{align}
Similarly, for the magnetothermal coefficient, we consider the in-plane thermal current density given by
\begin{align}
\label{eqellcur}
{\mathbf{\bar J}}^{\rm th,\chi}_s & = -\,
\tau  \int \frac{d^3 \mathbf{k} }  {(2 \, \pi)^3} \, \mathcal{D}^{\chi}_{\tilde s}  
\left (\xi^{\chi}_{\tilde s} - \mu{^{\chi}} \right )^2  
\left ( \boldsymbol{w}^{\chi}_{\tilde s} +  \boldsymbol{W}^{\chi}_{\tilde s} \right ) 
\left[  \left ( \boldsymbol{w}^{\chi}_{\tilde s} +  \boldsymbol{W}^{\chi}_{\tilde s}\right ) 
\cdot \frac{( - \,\nabla_{\mathbf{r}} T )}{T} \right ] 
f^\prime_0 (\xi^{\chi}_{\tilde s})\, ,
\end{align} 
which leads to the nonzero in-plane components expressed as
\begin{align}
\label{eqell0}
\left ( {\bar \ell}^{\chi}_{\tilde s} \right )_{ij}
=  - \,\tau   \int \frac{d^3 \mathbf{k} }  {(2 \, \pi)^3} \,
\mathcal{D}^{\chi}_{\tilde s} \, \frac{  (\xi^{\chi}_{\tilde s} - \mu{^{\chi}})^2}{T} 
\left[  (w^{\chi}_{\tilde s})_i +  (W^{\chi}_{\tilde s})_i \right ] 
\left[ (w^{\chi}_{\tilde s})_j +  (W^{\chi}_{\tilde s})_j \right ]
f^\prime_0 (\xi^{\chi}_{\tilde s})\,.
\end{align}

\section{Magnetoelectric conductivity} 
\label{secsigma}

The derivation of the various parts of the magnetoelectric conductivity tensor, as explained in Sec.~\ref{seclinear}, has been detailed in Appendix~\ref{appLMC}. Here, we will specifically focus only on the parts involving intranode-only scatterings.

For the intranode-only in-plane parts, we will consider the inclusion of the pseudomagnetic fields, in addition to the actual magnetic fields. This is because, we are interested in investigating how the strain can affect the linear response. For this purpose, we subject the sample to elastic deformations \cite{ips_rahul_ph_strain, ips-ruiz} such that the net effective magnetic field at a single node is given by
\begin{align}
\mathbf B^{\text{tot}} (\chi)= \mathbf B +  \chi\,\mathbf B_5\,,\quad
\mathbf B_5 = B_5  \left( \cos \theta_5 \,{\mathbf{\hat x}} 
+ \sin \theta_5 \, {\boldsymbol{\hat y}}\right),
\end{align} 
where $\mathbf {B }_5 $ is the emergent pseudomagnetic field due to strain (cf. Fig.~\ref{figsetup}). A pseudoelectric field $\mathbf E_5$, the counterpart of $\mathbf B_5$, can also be generated on dynamically stretching and compressing the crystal along an axis \cite{pikulin_gauge} (for instance, by driving longitudinal sound waves).
Then the net effective electric field is $\mathbf{E}^{\text{tot}} (\chi) = \mathbf E +  \chi\,\mathbf E_5$. We note that, while the physical electromagnetic fields couple to all the quasiparticles in the same way (irrespective of their chirality), the sign of the coupling of the pseudoelectromagnetic fields depends on $\chi$, which reflects their axial nature. For this part, we limit ourselves to terms upto $\order{|\mathbf B^{\text{tot}} |^ 3 }$.

Let us discuss the possibility of the presence of linear-in-$B$ terms in the diagonal components. When the system is subjected purely to homogeneous external fields (without any strain applied on the system), the Onsager-Casimir reciprocity relation~\cite{onsager31_reciprocal, onsager2, onsager3} for the diagonal components, viz. $(\bar \sigma^{\chi}_{\tilde s})_{ii} (\mathbf{B}) = (\bar \sigma^{\chi}_{\tilde s})_{ii} (-\mathbf{B})$ , is applicable --- it forbids any term in the LMC which has an odd power of $B$, unless the change of sign in $\mathbf B$ is compensated for by a change of sign in another parameter in the system \cite{cortijo16_linear, rahul-jpcm}. The pseudomagnetic field provides us with such a sign-compensating parameter, leading to
\begin{align} 
\label{eqonsager}
	(\bar \sigma^{\chi}_{\tilde s})_{ii} (\mathbf{B}, \mathbf{B}_5) = 
	(\bar \sigma^{\chi}_{\tilde s})_{ii} (-\mathbf{B}, -\mathbf{B}_5)\,,
\end{align}
thus fulfilling the Onsager-Casimir constraints. This suggests that, in the presence of a nonzero $\mathbf B_5$, a term linearly dependent on $B $ is possible.

The derivation of the non-anomalous-Hall contribution with intranode-only scatterings (without the Lorentz-force part) has been detailed in Appendices~\ref{appbcomm}.
To analyze the presence of various topological features, we divide up ${\bar \sigma}^{\chi}_{s}  $ into three parts as follows:
\begin{align}
\label{eqsig3parts}
{\bar \sigma}^{\chi}_{\tilde s}  =  
\sigma^{\chi, \text{Drude}}_{\tilde s} + \sigma^{\chi, \text{BC}}_{\tilde s} 
+ \sigma^{\chi, m}_{\tilde s} \,.
\end{align} 
Here, (1) the first part is the one which is independent of $ \mathbf B^{\text{tot}} $, also known as the Drude contribution;
(2) the second part arises solely due to the effect of the BC and survives when OMM is set to zero; and
(3) the third part is the one which goes to zero if OMM is ignored.

From the above expression, we now discuss the behaviour of the LMC and the PHC as functions of $B$ and $\theta$, the latter being the angle between $\mathbf{E}$ and $\mathbf{B}$. For plotting the nature of the in-plane components of the conductivity tensor, we define
\begin{align}
\label{eqSig}
& \Sigma_{ij} ( \mathbf B^{\text{tot}} )
 = {\bar \sigma}^+_{ij} - {\bar \sigma}^+_{ij} 
 \Big \vert_{\mathbf{B} =\mathbf B_5 = \boldsymbol{0} }\,,
\end{align}
where we have set $\chi= + 1$ for a charge-$ 4 $ RSW node. Furthermore, we have subtracted off the Drude contributions (which refer to the $ \mathbf B^{\text{tot}} $-independent parts), so that we can focus on the dependence controlled by applied magnetic (and pseudomagnetic) fields.
The ranges of the values of the parameters in some realistic scenarios have been shown in Table \ref{table_params}, which we have used in our plots.

\begin{figure}[t]
\centering
{\includegraphics[width= 0.65 \textwidth]{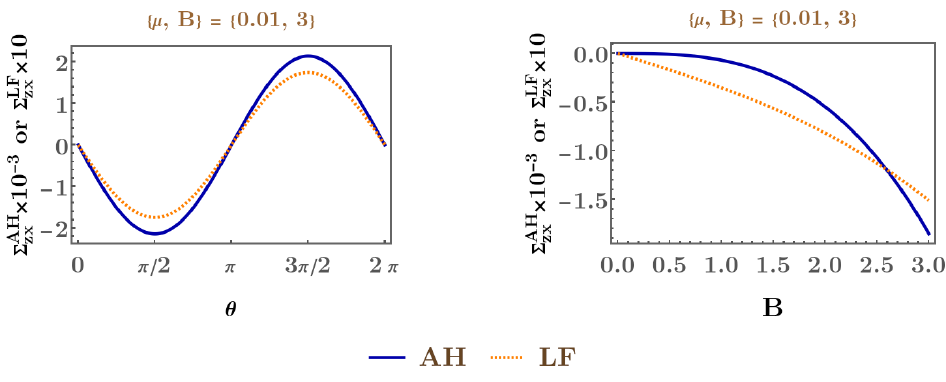}}
\caption{The plots show the variation of the transverse $zx$-components of the magnetoelectric conductivity (in the units of eV) with the angle $\theta$ (left panel) and the magnitude $B$ in the units of eV$^{2}$ (right panel), after setting $\mathbf B_5 = \mathbf 0$. Along the vertical axis, we have plotted the anomalous-Hall-only (denoted by ``AH'') and the Lorentz-force-only (denoted by ``LF'') parts, with the colour-coding shown in the plotlegends. The values of the fixed parameters are indicated in each plotlabel. For obtaining all the curves, we have set $v_0 = 0.005$, $\tau=151$ eV$^{-1}$, and $T = 10^{-3}$ eV.
\label{figszx}}
\end{figure}

\subsection{Out-of-plane Hall components}
\label{secah_lf}

In this subsection, we set $\mathbf B_5 = \mathbf 0 $, because our main aim is to investigate the nature of the conductivity in setups considered in Ref.~\cite{claudia-multifold}, where strain-induced pseudomagnetic fields are not considered. As discussed in Sec~\ref{seclinear}, nonzero out-of-plane components are generated only from the anomalous-Hall and Lorentz-force parts.

For the intrinsic anomalous-Hall part, following the treatment in Appendix~\ref{appsigmaAH},
the application of the Sommerfeld expansion yields
\begin{align}
\label{eqah}
(\sigma^{ {\text{AH}}}_{\tilde s})_{zx}  & =   -\,
 \frac{e^3 \, {\tilde s} \, v_0 \, \mathcal{G}_{\tilde s} \, B_y}
{60 \, \pi^2 }  
   \int 
\frac{d\epsilon}{\epsilon} \left [ 10  \, \epsilon^2 \, f^\prime_0 (\varepsilon_{\tilde s} ) 
+ e^2 \,  {\tilde s}^2 \, v_0^4 \, 
\mathcal{G}_{\tilde s}^2  \, B^2 \, f^{\prime \prime \prime}_0 (\varepsilon_{\tilde s} ) \right ] 
\nn & = -\,
\frac{ e^3 \, {\tilde s} \, v_0 \, \mathcal{G}_{\tilde s} \, B_y }
	 {30 \, \pi^2}
\left [ 5 \, \Upsilon_{-1}(\mu ,T) + 6 \, B^2 \, 
 e^2 \,  {\tilde s}^2 \, v_0^4 \, \mathcal{G}_{\tilde s}^2 \, \Upsilon_{-5}(\mu ,T) \right ], 
\nn	\Upsilon_n(\mu , T) & =  
	\mu^n \left [ 1 + \frac{\pi^2 \, T^2 \, n \, (n - 1)}
	{6  \, \mu^2} + \ldots   \right  ].
\end{align}
We have removed the $\chi$-superscript here because, in the absence of $\mathbf B_5 $, the above expression is independent of $\chi$.

\begin{figure}[t]
\centering
{\includegraphics[width= 0.65 \textwidth]{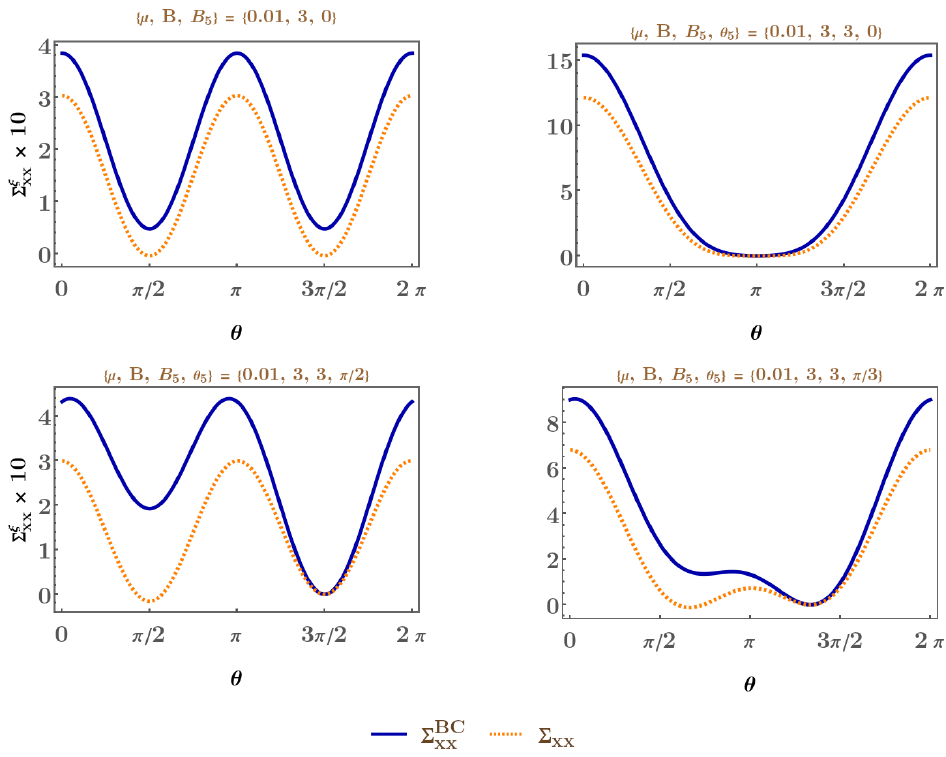}}
\caption{The plots show the variation of the total LMC (in the units of eV) with the angle $\theta$ between $\mathbf B$ and $\mathbf E $, after subtracting off the Drude part. The superscript ``BC'' indicates that, for those curves, the OMM contributions have been set to zero. We have used the superscript $\xi$ to indicate that, along the vertical axis, we have plotted the BC-only and OMM-added parts, with the colour-coding shown in the plotlegends. Here, we have included the effects of strain in the form $\mathbf B_5$. The values of $B$ and $B_5$ (in  eV$^{2}$) are indicated in each plotlabel, along with the value of $\theta_5 $. For obtaining all the curves, we have set $\mu =0.01 $ eV, $v_0 = 0.005$, $\tau=151$ eV$^{-1}$, and $T = 10^{-3} $ eV.
\label{figsxx1}}
\end{figure}

The Lorentz-force contribution, derived in Appendix~\ref{applor}, gives rise to only a nonzero $zx$-component for ${\sigma}^{\chi, \rm LF}_{\tilde s}$. Using Eq.~\eqref{eqsigLF}, the final expression is captured by
\begin{align} 
\label{eqsigmazx}
\left (\sigma^{\rm LF}_{\tilde s} \right )_{zx}  &= 
-\, \frac{e^3 \, {\tilde s} \, \tau^2 \, v_0} {30\, \pi^2} \,
B_y  \left[ 5 \, \Upsilon_1(\mu ,T) 
+ e^2 \, {\tilde s}^2 \, v_0^4 \left(\mathcal{G}_{\tilde s}^2 
- 8 \, \mathcal{G}_{\tilde s} \, {\tilde s}^2 + 3 \, {\tilde s}^4  \right) 
B^2 \, \Upsilon_{-3}(\mu ,T)\right]. 
\end{align}
Again, we have removed the $\chi$-superscript here because, in the absence of $\mathbf B_5 $, the conductivity-expression is independent of $\chi$.

We find a couple of similarities between the intrinsic anomalous-Hall part and the Lorentz-force contribution: they both have only odd powers of $B$, and only their out-of-plane Hall components survive. In Fig.~\ref{figszx}, we have shown the nature of
\begin{align}
\Sigma^{\rm AH}_{zx} = \sum_{\tilde s } \left (\sigma^{\rm AF}_{\tilde s} \right )_{zx} 
\text{ and }
\Sigma^{\rm LF}_{zx} = \sum_{\tilde s } \left (\sigma^{\rm LF}_{\tilde s} \right )_{zx} \,,
\end{align}
by choosing some representative parameter values. 
In a recent experimental investigation \cite{claudia-multifold}, the authors have showed the importance of the cubic-in-$B$ terms for multifold semimetals, arising in $\left (\sigma^{\rm AH}_{s} \right )_{zx}$ and $\left (\sigma^{\rm LF}_{s} \right )_{zx}$. Although they have studied fourfold band-crossings, they are restricted to doubled pseudospin-1/2 quasiparticles (i.e., two degenerate copies of a Weyl node with the same monopole charge). Hence, their result is obtained simply by multiplying the result for a WSM with a factor of 2 (and another factor of 2 if we want to account for the spin-degeneracy for materials with negligible spin-orbit coupling).
They have fitted their data using a semiclassical Boltzmann theory, similar to our treatment. Their findings demonstrate that a negative magnetoresistance originates from the chiral anomaly, despite a sizable and detrimental OMM contribution, which was previously unaccounted for. Our calculations take into account the nontrivial fourfold band-crossing of the RSW case, which is not simply two copies of a Weyl node.

\subsection{Longitudinal magnetoelectric conductivity (LMC)} 

\begin{figure}[t]
\centering
{\includegraphics[width= 0.65 \textwidth]{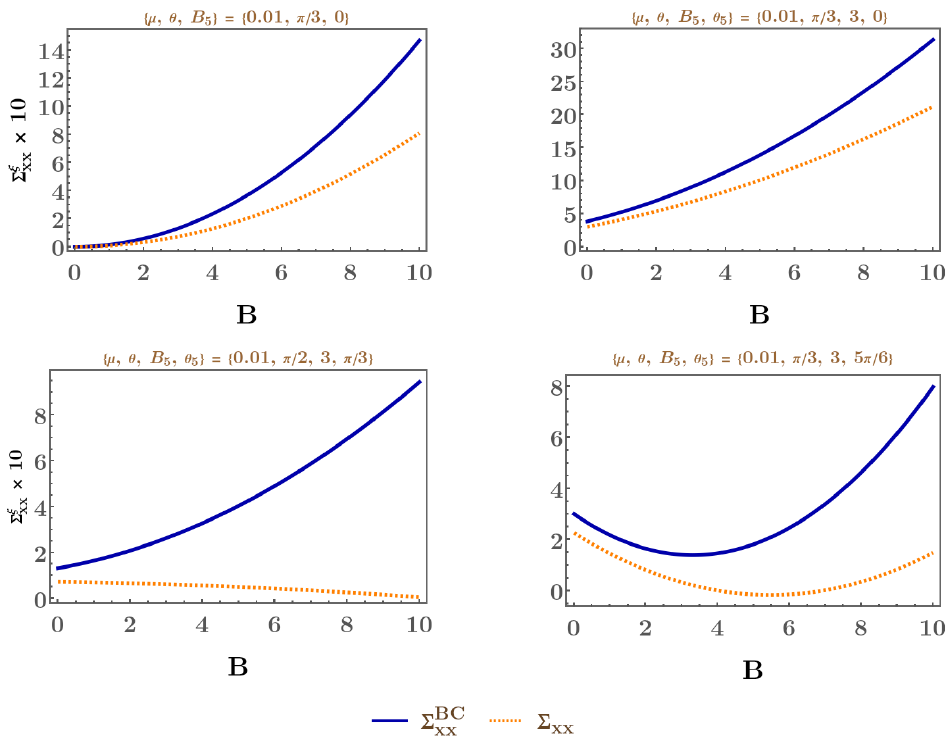}}
\caption{
The plots show the variation of the total LMC (in the units of eV) with the magnitude $B$ of the physical magnetic field, after subtracting off the Drude part. The superscript ``BC'' indicates that, for those curves, the OMM contributions have been set to zero. We have used the superscript $\xi$ to indicate that, along the vertical axis, we have plotted the BC-only and OMM-added parts, with the colour-coding shown in the plotlegends.
Here, we have included the effects of strain in the form $\mathbf B_5$. The value of $B_5$ (in  eV$^{2}$) is indicated in each plotlabel, along with the values of $\theta $ and $\theta_5 $. For obtaining all the curves, we have set $\mu =0.01 $ eV, $v_0 = 0.005$, $\tau=151$ eV$^{-1}$, and $T = 10^{-3} $ eV.
\label{figsxx2}}
\end{figure}

Using the expressions derived in Appendix~\ref{appbcomm}, the longitudinal (or diagonal) in-plane component is given by
\begin{align}
\left( {\bar \sigma}^{\chi}_{\tilde s} \right)_{xx}  =  
\left( \sigma^{\chi, \text{Drude}}_{\tilde s}\right)_{xx} + 
\left( \sigma^{\chi, \text{BC}}_{\tilde s} \right)_{xx}
+ \left( \sigma^{\chi, m}_{\tilde s} \right)_{xx}\,,
\end{align} 
where
\begin{align}
& \left( \sigma^{\chi, \text{Drude}}_{\tilde s} \right)_{xx} 
 =  \frac{e^2 \, \tau }
 {6 \,\pi^2 \, \tilde s \, v_0} \, \Upsilon_2( \mu ,T) \,,\quad
 \left( \sigma^{\chi, \text{BC}}_{\tilde s} \right)_{xx}
 =    \frac{e^4 \, \tau \,  {\tilde s}^5 \,  v_0^3} 
{30 \, \pi^2 }   
\left [  8  \,   (B^{\text{tot}}_x)^2 +   (B^{\text{tot}}_y)^2 \right ]
 \Upsilon_{-2}( \mu ,T) \,,\nn
& \left( \sigma^{\chi, m}_{\tilde s} \right)_{xx}  = 
 \frac{e^4 \, \tau \,  \tilde s \,  v_0^3 \, \mathcal{G}_{\tilde s}} 
 {30 \, \pi^2 }   
 \left [ \left( - \, 9 \, \tilde s^2 +5 \, \mathcal{G}_{\tilde s} \right) 
 (B^{\text{tot}}_x)^2 -3 \, \tilde s^2  
 \,  (B^{\text{tot}}_y)^2 \right ] \Upsilon_{-2}( \mu ,T)  \,.
\end{align}
Adding up the three parts, the total gives us
\begin{align} 
\label{eqsigxx}
\left( {\bar \sigma}^{\chi}_{\tilde s} \right)_{xx} 
=  \frac{e^2 \, \tau }{6 \, \pi^2 \, \tilde s\, v_0} \, \Upsilon_2( \mu ,T) 
+   \frac{e^4 \, \tau \,  {\tilde s} \,  v_0^3}
{30 \, \pi^2 }   
 \left[ 
(B^{\text{tot}}_x)^2
 \left(8 \, {\tilde s}^4 -9 \,{\tilde s}^2\, \mathcal{G}_{\tilde s}
 +5 \,\mathcal{G}_{\tilde s}^2 \right)
 +
 {\tilde s}^2 \, (B^{\text{tot}}_y)^2 
\left( {\tilde s}^2-3 \, \mathcal{G}_{\tilde s} \right)
 \right ] \Upsilon_{-2}( \mu ,T) \, .
\end{align}

Let us define the functions
\begin{align}
f_x^{\rm BC} (\tilde s) = 8\,\tilde s^5\,,\quad
f_x^{m}(\tilde s) = \tilde s \,  \mathcal{G}_{\tilde s}
 \left( - \, 9 \, \tilde s^2 +5 \, \mathcal{G}_{\tilde s} \right) \,, \quad
f_y^{\rm BC}(\tilde s) =\tilde s^5\,,\quad
f_y^{m}(\tilde s) = -\, 3\,{\tilde s}^3 \, \mathcal{G}_{\tilde s}  \,,
\end{align}
which are the coefficients of the $(B^{\text{tot}}_x)^2$ and $(B^{\text{tot}}_y)^2$ terms, when we consider the BC-only and OMM-effects separately. The supercripts and the subscrtipts indicate which part of the response and which component of $\mathbf B^{\text{tot}} $ they are referring to.
We find that $f_x^{\rm BC} (1/2) = 0.25 $, $f_x^{m}(1/2) = 5.6875 $, $f_x^{\rm BC} (3/2) = 60.75 $, $f_x^{m}(3/2) = - \,18.5625$, $\sum _{\tilde s} f_x^{\rm BC} (\tilde s) = 61 $, $\sum _{\tilde s} f_x^{m} (\tilde s) = -\, 12.875 $,
$f_y^{\rm BC} (1/2) = 0.03125 $, $f_y^{m}(1/2) = - 1.3125 $, $f_y^{\rm BC} (3/2) = 7.59375 $, $f_y^{m}(3/2) = - \,5.0625 $,
$\sum _{\tilde s} f_y^{\rm BC} (\tilde s) = 7.625 $, and $\sum _{\tilde s} f_y^{m} (\tilde s) = -\, 6.375 $.
These results tell us that
\begin{enumerate}
\item $B_{x}^{\text{tot}}$-part: (a) for $\tilde s =1/2$, the OMM-part adds up to the BC-only term, thus increasing the overall response; (b) for $\tilde s = 3/2$, the OMM-part acts in opposition to the BC-only term, thus decreasing the overall response. However, after we sum over the two bands, the contribution from $\tilde s = 3/2$ dominates, leading to an overall detrimental effect of nonzero OMM, compared to the scenario when we ignore it.

\item $B_{y}^{\text{tot}}$-part: For each of $\tilde s =1/2$ and $\tilde s = 3/2$, the OMM-part acts in opposition to the BC-only term, thus decreasing the overall response. Hence, there is an overall detrimental effect of nonzero OMM compared to the scenario when we ignore it.

\end{enumerate}

We have illustrated the behaviour of $ \Sigma_{xx}^{\rm BC} $ and $ \Sigma_{xx} $ in Figs.~\ref{figsxx1} and \ref{figsxx2}, where the superscript ``BC'' indicates that the contributions come from the BC-only parts. While the curves in Fig.~\ref{figsxx1} show the variation of the response as a function of $\theta $, those in Fig.~\ref{figsxx2} capture the dependence on $ B $. In agreement with our comparison of the $ f $-values, we find that a nonzero OMM always reduces the response. In the first subfigure of Fig.~\ref{figsxx1}, we have the curves with $\mathbf B_5 = \mathbf 0 $. Comparing it with the remaining subfigures, we find that a nonzero $\mathbf B_5$-part changes the periodicity with respect to $\theta $ from $\pi $ to $2 \pi $, which results from the emergence of terms linearly proportional to $ B $, rather than just the $\propto B^2 $ ones (see Refs.~\cite{ips_rahul_ph_strain, ips-ruiz} for a similar behaviour in Weyl and multi-Weyl semimetals).

\begin{figure}[t]
\centering
\includegraphics[width= 0.65 \textwidth]{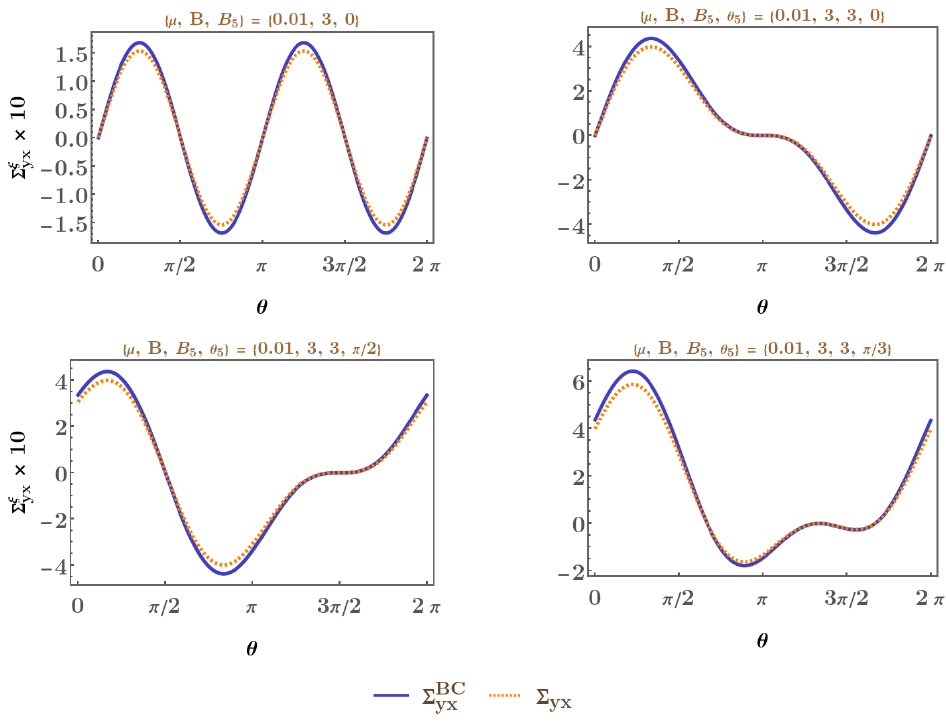}
\caption{The plots show the variation of the total PHC (in the units of eV) with the angle $\theta$ between $\mathbf B$ and $\mathbf E $, after subtracting off the Drude part. The superscript ``BC'' indicates that, for those curves, the OMM contributions have been set to zero. We have used the superscript $\xi$ to indicate that, along the vertical axis, we have plotted the BC-only and OMM-added parts, with the colour-coding shown in the plotlegends. Here, we have included the effects of strain in the form $\mathbf B_5$. The values of $B$ and $B_5$ (in  eV$^{2}$) are indicated in each plotlabel, along with the value of $\theta_5 $. For obtaining all the curves, we have set $\mu =0.01 $ eV, $v_0 = 0.005$, $\tau=151$ eV$^{-1}$, and $T = 10^{-3} $ eV.
\label{figsyx1}}
\end{figure}

\begin{figure}[t]
\centering
\includegraphics[width= 0.65 \textwidth]{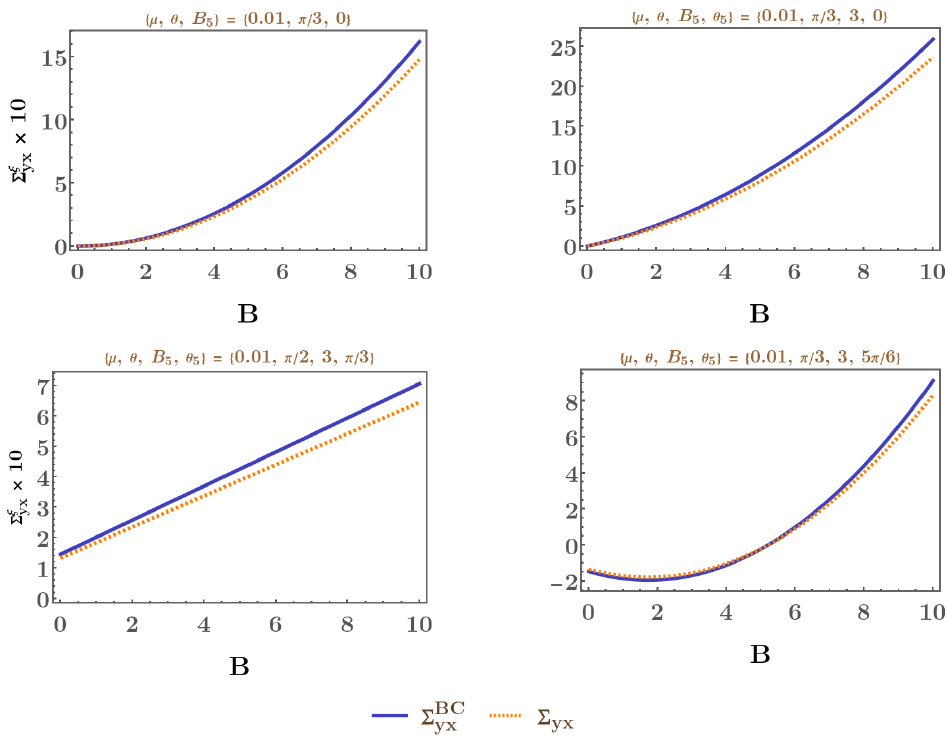}
\caption{
The plots show the variation of the total PHC (in the units of eV) with the magnitude $B$ of the physical magnetic field, after subtracting off the Drude part. The superscript ``BC'' indicates that, for those curves, the OMM contributions have been set to zero. We have used the superscript $\xi$ to indicate that, along the vertical axis, we have plotted the BC-only and OMM-added parts, with the colour-coding shown in the plotlegends.
Here, we have included the effects of strain in the form $\mathbf B_5$. The value of $B_5$ (in  eV$^{2}$) is indicated in each plotlabel, along with the values of $\theta $ and $\theta_5 $. For obtaining all the curves, we have set $\mu =0.01 $ eV, $v_0 = 0.005$, $\tau=151$ eV$^{-1}$, and $T = 10^{-3} $ eV.
\label{figsyx2}}
\end{figure}

\subsection{Transverse magnetoelectric conductivity (PHC)}

Using the expressions derived in Appendix~\ref{appbcomm}, the transverse in-plane component is given by
\begin{align}
\left( {\bar \sigma}^{\chi}_{\tilde s} \right)_{yx}  =  
\left( \sigma^{\chi, \text{Drude}}_{\tilde s}\right)_{yx} + 
\left( \sigma^{\chi, \text{BC}}_{\tilde s} \right)_{yx}
+ \left( \sigma^{\chi, m}_{\tilde s} \right)_{yx}\,.
\end{align} 
where
\begin{align}
& \left( \sigma^{\chi, \text{Drude}}_{\tilde s}\right)_{yx} = 0\,, \quad
\left( \sigma^{\chi, \text{BC}}_{\tilde s} \right)_{yx} 
= \frac{7 \,  e^4 \, \tau  \,  {\tilde s}^5  \, v_0^3}
{30 \, \pi^2} \, B^{\text{tot}}_x \, B^{\text{tot}}_y \, 
\Upsilon_{-2}( \mu ,T)\,, \nn
& \left( \sigma^{\chi, m}_{\tilde s} \right)_{yx} = 
\frac{e^4 \, \tau \, {\tilde s} \,  v_0^3 \,\mathcal{G}_{\tilde s} }
{30 \, \pi^2} \, B^{\text{tot}}_x \, B^{\text{tot}}_y 
 \left( - \, 6 \, {\tilde s}^2  + 5  \,\mathcal{G}_{\tilde s} \, \right) 
  \Upsilon_{-2}( \mu ,T)\,.
\end{align}
The addition of the two nonzero parts gives us the planar Hall conductivity (PHC) as
\begin{align}
\left( {\bar \sigma}^{\chi}_{\tilde s} \right)_{yx} = 
\frac{e^4 \, \tau \, {\tilde s} \, v_0^3 } {30 \, \pi^2} \, 
B^{\text{tot}}_x \, B^{\text{tot}}_y  
\left(7 \, {\tilde s}^4  -  6 \,{\tilde s}^2 \,\mathcal{G}_{\tilde s} 
+ 5  \,\mathcal{G}_{\tilde s}^2 \, \right)  \Upsilon_{-2}( \mu ,T)\,.
\end{align}

Let us define the functions
\begin{align}
g^{\rm BC} (\tilde s) = 7\,\tilde s^5\,,\quad
g^{m}(\tilde s) = \tilde s \,  \mathcal{G}_{\tilde s}
 \left( - \, 6 \, \tilde s^2 +5 \, \mathcal{G}_{\tilde s} \right)  \,,
\end{align}
which are the coefficients of the BC-only and OMM-effects separately. The supercripts indicate which part they are referring to.
We find that $ g^{\rm BC} (1/2) = 0.21875 $, $ g^{m}(1/2) = 6.34375 $, $ g^{\rm BC} (3/2) = 53.1563 $, $ g^{m}(3/2) = - \,10.9688 $, $\sum_{\tilde s} g^{\rm BC} (\tilde s) = 53.375 $, and $\sum _{\tilde s} g^{m} (\tilde s) = -\, 4.625 $.
These results tell us that (1) for $\tilde s =1/2$, the OMM-part adds up to the BC-only term, thus increasing the overall response; (2) for $\tilde s = 3/2$, the OMM-part acts in opposition to the BC-only term, thus decreasing the overall response. However, after we sum over the two bands, the contribution of the $\tilde s = 3/2$-band dominates, leading to an overall detrimental effect of nonzero OMM, compared to the scenario when we ignore it.

We have illustrated the behaviour of $ \Sigma_{yx}^{\rm BC} $ and $ \Sigma_{yx} $ in Figs.~\ref{figsyx1} and \ref{figsyx2}, where the superscript ``BC'' indicates that the contributions come from the BC-only parts. While the curves in Fig.~\ref{figsyx1} show the variation of the response as a function of $\theta $, those in Fig.~\ref{figsyx2} capture the dependence on $ B $. In agreement with our comparison of the $ g $-values, we find that a nonzero OMM always reduces the response. In the first subfigure of Fig.~\ref{figsyx1}, we have the curves with $\mathbf B_5 = \mathbf 0 $. Comparing it with the remaining subfigures, we find that a nonzero $\mathbf B_5$-part changes the periodicity with respect to $\theta $ from $\pi $ to $2 \pi $, which results from the emergence of terms linearly proportional to $ B $, rather than just the $\propto B^2 $ ones (see Refs.~\cite{ips_rahul_ph_strain, ips-ruiz} for a similar behaviour in Weyl and multi-Weyl semimetals).

\section{Magnetothermoelectric conductivity and magnetothermal coefficient}
 \label{secalpha}

We divide up the expressions for ${\bar \alpha}^{\chi}_{\tilde s}  $ and ${\bar \ell}^{\chi}_{\tilde s}  $, shown in Eqs.~\eqref{eqalpha} and \eqref{eqell0}, into three parts as
\begin{align}
\label{eqalpha3parts}
{\bar \alpha}^{\chi}_{\tilde s}  =  
\alpha^{\chi, \text{Drude}}_{\tilde s} + \alpha^{\chi, \text{BC}}_{\tilde s} 
+ \alpha^{\chi, m}_{\tilde s} \text{ and }
{\bar \ell}^{\chi}_{\tilde s}  =  
\ell^{\chi, \text{Drude}}_{\tilde s} + \ell^{\chi, \text{BC}}_{\tilde s} 
+ \ell^{\chi, m}_{\tilde s} \,.
\end{align} 
Analogous to the case of ${\bar \sigma}^{\chi}_{\tilde s}$, (1) the first part stands for the Drude contribution; (2) the second part arises solely due to the effect of the BC and survives when OMM is set to zero; and (3) the third part is the one which goes to zero if OMM is ignored.

\subsection{Magnetothermoelectric conductivity}

Using the expressions derived in Appendix~\ref{appalpha}, the longitudinal (or diagonal) in-plane component of $ {\bar \alpha}^{\chi}_{\tilde s} $, also known as the longitudinal thermoelectric coefficient (LTEC), is given by
\begin{align}
\left( {\bar \alpha}^{\chi}_{\tilde s} \right)_{xx}  =  
\left( \alpha^{\chi, \text{Drude}}_{\tilde s}\right)_{xx} + 
\left( \alpha^{\chi, \text{BC}}_{\tilde s} \right)_{xx}
+ \left( \alpha^{\chi, m}_{\tilde s} \right)_{xx}\,,
\end{align} 
where
\begin{align}
&	\left( \alpha^{\chi, \text{Drude}}_{\tilde s}\right)_{xx} = 
- \,\frac{e  \,\tau \,  \mu \, T}{9 \, {\tilde s} \, v_0}\,,
\quad	\left( \alpha^{\chi, \text{BC}}_{\tilde s} \right)_{xx} = 
	\frac{e^3 \, \tau \,  {\tilde s}^5 \,  v_0^3 \, T}
	{45 \, \mu^3 }   
\left [ 8 \, (B^{\text{tot}}_x)^2  +  (B^{\text{tot}}_y)^2 \right ] ,
\nn &	
\left( \alpha^{\chi, m}_{\tilde s} \right)_{xx}
= \frac{e^3 \, \tau \,  {\tilde s} \,  v_0^3 \, \mathcal{G}_{\tilde s} \, T}
{45 \, \mu^3 }  
\left [ (B^{\text{tot}}_x)^2 
\left( -9 \, {\tilde s}^2 +5 \, \mathcal{G}_{\tilde s} \right) 
- 3 \, {\tilde s}^2 \, (B^{\text{tot}}_y)^2 \right ] .
\end{align}
The total expression for the LTEC reads
\begin{align}
\left( {\bar \alpha}^{\chi}_{\tilde s} \right)_{xx}   &= -\,
	\frac{e  \,\tau \, \mu \, T}{9 \, {\tilde s}  \, v_0} +
	 \frac{e^3 \, \tau \,  {\tilde s}  \,  v_0^3 \, T}
	{45 \,  \mu^3 }  
\left [ (B^{\text{tot}}_x)^2 
\left(8 \,{\tilde s}^4 
- 9 \, {\tilde s} ^2 \, \mathcal{G}_{\tilde s}
+ 5 \, \mathcal{G}_{\tilde s}^2\right) 
+ {\tilde s}^2 \, (B^{\text{tot}}_y)^2 
\left( {\tilde s}^2-3 \, \mathcal{G}_{\tilde s}\right) \right ] .
\end{align}

Again, the expressions derived in Appendix~\ref{appalpha} lead to the following expression for the transverse in-plane component of $ {\bar \alpha}^{\chi}_{\tilde s} $, also known as the transverse thermoelectric coefficient (TTEC):
\begin{align}
\left( {\bar \alpha}^{\chi}_{\tilde s} \right)_{yx}  =  
\left( \alpha^{\chi, \text{Drude}}_{\tilde s}\right)_{yx} + 
\left( \alpha^{\chi, \text{BC}}_{\tilde s} \right)_{yx}
+ \left( \alpha^{\chi, m}_{\tilde s} \right)_{yx}\,,
\end{align} 
where
\begin{align}
&	\left( \alpha^{\chi, \text{Drude}}_{\tilde s}\right)_{yx} = 
0\,,
\quad	\left( \alpha^{\chi, \text{BC}}_{\tilde s} \right)_{yx} = 
\frac{7 \, e^3 \, \tau \, {\tilde s}^5 \, v_0^3  \, T}
{45 \, \mu^3}  \, B^{\text{tot}}_x \, B^{\text{tot}}_y\, ,
\quad	
\left( \alpha^{\chi, m}_{\tilde s} \right)_{yx}
=  \frac{e^3 \, \tau \, \tilde s \, v_0^3  \, T}
{45 \,  \mu^3}  \, 
B^{\text{tot}}_x \, B^{\text{tot}}_y 
\left(-6 \, {\tilde s}^2 \,\mathcal{G}_{\tilde s}
+ 5 \, \mathcal{G}_{\tilde s}^2\right) .
\end{align}
The total expression for the TTEC reads
\begin{align}
\left( {\bar \alpha}^{\chi}_{\tilde s} \right)_{xx} =
	\frac{e^3 \, \tau \, {\tilde s} \, v_0^3 \, T}
	{45 \, \mu^3}  \, B^{\text{tot}}_x 
	\, B^{\text{tot}}_y \left(7 \, {\tilde s}^4 
-6 \, {\tilde s}^2 \, \mathcal{G}_{\tilde s}
+ 5 \, \mathcal{G}_{\tilde s}^2 \right) .
\end{align}

 \subsection{Magnetothermal coefficient}

Using the expressions derived in Appendix~\ref{appell}, the longitudinal (or diagonal) in-plane component of $ {\bar \ell}^{\chi}_{\tilde s} $ is given by
\begin{align}
\left( {\bar \ell}^{\chi}_{\tilde s} \right)_{xx}  =  
\left( \ell^{\chi, \text{Drude}}_{\tilde s}\right)_{xx} + 
\left( \ell^{\chi, \text{BC}}_{\tilde s} \right)_{xx}
+ \left( \ell^{\chi, m}_{\tilde s} \right)_{xx}\,,
\end{align} 
where
\begin{align}
	& \left( \ell^{\chi, \text{Drude}}_{\tilde s}\right)_{xx} = 
 \frac{ \mu^2 \, \tau \, T}
 {18 \, \tilde{s} \, v_0} \,,\quad
	\left( \ell^{\chi, \text{BC}}_{\tilde s} \right)_{xx} =   
	\frac{e^2 \, s^5 \, \tau \,v_0^3 \, T}
	{90 \,  \mu^2 }
	\
	\left [ 8 \,(B^{\text{tot}}_x)^2 + (B^{\text{tot}}_y)^2 \right ]\,, \nn
	&(\ell^{\chi, m}_{\tilde s})_{xx} =  \frac{e^2 \, \tilde{s} \, \tau \, v_0^3 \, \mathcal{G}_{s} \, T }
{90 \,  \mu^2 } \left [ (B^{\text{tot}}_x)^2 
	\left( 5 \, \mathcal{G}_{\tilde s} -9 \, {\tilde s}^2  \right) 
	- 3 \, {\tilde s}^2 \, (B^{\text{tot}}_y)^2 \right ].
\end{align}
The sum of all the parts reads
\begin{align}
\left( {\bar \ell}^{\chi}_{\tilde s} \right)_{xx}  & =  
\frac{e^2 \, {\tilde s} \,\tau \,v_0^3 \, T}
{90  \,  \mu^2 } \;
\bigg [ \mathcal{G}_{\tilde{s}}  \left \{ (B^{\text{tot}}_x)^2 
\left( 5 \, \mathcal{G}_{\tilde s} -9 \, {\tilde s}^2  \right) 
	- 3 \, {\tilde s}^2 \left (B^{\text{tot}}_y \right )^2 
\right \}
+ {\tilde s}^4   \left \lbrace 8  \left (B^{\text{tot}}_x \right )^2
+ (B^{\text{tot}}_y   )^2 \right \rbrace
	\bigg ] \
+ \frac{ \mu^2 \, \tau \, T}
	{18 \, \tilde{s} \, v_0} \,.
\end{align}

Again, the expressions derived in Appendix~\ref{appell} lead to the following expression for the transverse in-plane component of $ {\bar \ell}^{\chi}_{\tilde s} $:
\begin{align}
\left( {\bar \ell}^{\chi}_{\tilde s} \right)_{yx}  =  
\left( \ell^{\chi, \text{Drude}}_{\tilde s}\right)_{yx} + 
\left( \ell^{\chi, \text{BC}}_{\tilde s} \right)_{yx}
+ \left( \ell^{\chi, m}_{\tilde s} \right)_{yx}\,,
\end{align} 
where
\begin{align}
& \left( \ell^{\chi, \text{Drude}}_{\tilde s}\right)_{yx} =  0 \,,\quad
\left( \ell^{\chi, \text{BC}}_{\tilde s} \right)_{yx} =  
\frac{7 \, e^2 \, {\tilde s}^5 \, \tau \,v_0^3 \, T}
{90 \,  \mu^2 } \,B^{\text{tot}}_x \, 
B^{\text{tot}}_y  \,,\quad
\left( \ell^{\chi, m}_{\tilde s} \right)_{yx} = 
\frac{e^2 \, \tilde{s} \, \tau \, v_0^3 \, \mathcal{G}_{s} \, T }
{90 \,   \mu^2 } B^{\text{tot}}_x B^{\text{tot}}_y 
\left( 5 \, \mathcal{G}_{\tilde s} -6 \, {\tilde s}^2 \,
\right) .
\end{align}
The sum of all the parts reads
\begin{align}
\left( {\bar \ell}^{\chi}_{\tilde s} \right)_{yx} =   
\frac{e^2 \, {\tilde s} \, \tau \,v_0^3 \, T } 
{ 90 \,  \mu^2 } \,B^{\text{tot}}_x \, B^{\text{tot}}_y 
\left[  7 \, \tilde{s}^4  + \mathcal{G}_{s} \, \left( 5 \, \mathcal{G}_{\tilde s} -6 \, {\tilde s}^2 
\right) \right ]  .
\end{align}

\subsection{Mott relation and Wiedemann-Franz law}

From the explicit expressions of $\bar \sigma^{\chi}_{\tilde s} $ and $\bar \alpha^{\chi}_{\tilde s} $ that we have demonstrated, we can immediately spot the relation
\begin{align}
\partial_{ \mu }  (\bar \sigma^{\chi}_{\tilde s}  )_{ij} 
= - \frac  {3\, e }  {\pi^2 \, T} \left (\bar \alpha^{\chi}_{\tilde s} \right )_{ij} 
+ \order{T^{2}}
\label{eqmott}
\end{align}
being satisfied. This is equivalent to satisfying the Mott relation, which holds in the limit $T \rightarrow 0 $ \cite{mermin}. In particular, we find that the Mott relation continues to hold in the presence of OMM, agreeing with the results of Ref.~\cite{xiao06_berry}, where generic settings for the linear response have been considered. 
From the explicit expressions of $\sigma^{\chi}_{\tilde s} $ and $\ell^{\chi}_{\tilde s} $, we find another relation, namely,
\begin{align}
( \bar \sigma^{\chi}_{\tilde s})_{ij} =
\frac{ 3 \,e^2 } {\pi^2\, T}   
\left (\bar \ell^{\chi}_{\tilde s} \right )_{ij}   + \order{T^{2}} ,
\label{eqwf}
\end{align}
being satisfied. This is equivalent to satisfying the Wiedemann-Franz law, which again holds in the limit $T \rightarrow 0 $ \cite{mermin}. Therefore, we find that the Wiedemann-Franz law also continues to be valid in the presence of OMM. 
Due to the Mott relation and the Wiedemann-Franz law, the behaviour of $ (\bar \alpha^{\chi}_{\tilde s})_{ij}  $ and $ \left (\bar \ell^{\chi}_{\tilde s} \right )_{ij}$ can be readily inferred from that of $(\bar \sigma^{\chi}_{\tilde s})_{ij} $. Hence, we do not provide separate plots and discussions for the $\bar \alpha^{\chi}_{\tilde s}$ and $\bar \ell^{\chi}_{\tilde s}$ tensors.

\section{Summary and future perspectives} 
\label{sec_summary}

In this paper, we have considered planar Hall and planar thermal Hall setups, where an RSW semimetal is subjected to the combined effects of an electric field $\mathbf E $ and/or temperature gradient $\nabla_{\mathbf r } T$. The $\mathbf E $ and $\nabla_{\mathbf r } T$ fields are assumed to be along the same direction. Since we have considered an isotropic RSW material, without any rotational-symmetry-breaking term (e.g., the tilting of the nodes), the plane in which the fields are applied makes no difference. For computing the in-plane components of the response tensors, we have added an elastic deformation, which gives rise to a chirality-dependent effective magnetic field $\mathbf B^{\text{tot}}$ consisting of two parts --- (1) the physical magnetic field $\mathbf B $ and (2) an emergent axial magnetic field $\mathbf B_5 $. This is captured by defining $ \mathbf B^{\text{tot}} = 
\mathbf B  + \chi \, \mathbf B_5 $ for the corresponding nodal point. Due to the chiral nature of $\mathbf B_5 $, its presence makes it possible to have linear-in-$B$ terms in the linear-response coefficients, which otherwise is ruled out in accordance with the Onsager-Casimir reciprocity relations. In all our calculations, the effects of the nontrivial topology of the RSW bandstructure have been captured through the inclusion of both the BC and the OMM. In many earlier works, the response in such nodal-point semimetals have been computed neglecting the OMM parts. However, following the treatment of some recent papers \cite{timm, das-agarwal_omm, onofre, ips-ruiz}, here we have included the OMM terms in a systematic way, and have emphasized on the importance of the consequence of a nonzero OMM, which anyway arises at the same footing as the BC.

The RSW semimetals provide a richer structure for obtaining the linear-response coefficients, compared to the WSMs, because of the fact that the former consists of four bands (rather than just two). Although each band still shows a linear-in-momentum dispersion, just like a WSM node, the constant of proportionality with $k$ changes from band to band. Furthermore, needless to say, the bands have differing BC and OMM, which then provide unequal contributions to the net response. We have clearly pointed out these aspects in our explicit derivations of the electric conductivity, thermoelectric conductivity, and thermal coefficient tensors, in the presence of a non-quantizing magnetic field. In particular, we have found that the OMM-contributed terms may oppose or add up to the BC-only parts, depending on which band we are considering.

Last, but not the least, we have determined the out-of-plane response comprising the intrinsic anomalous-Hall and the Lorentz-force-contributed currents. These terms inherently consist of only odd powers of $B$, giving rise to linear-in-$B$ and cubic-in-$B $ dependence, when we limit ourselves to expanding the expressions upto order $B^3$. These terms corroborate the findings of some recent experimental results \cite{claudia-multifold}, which have found clear signatures of the importance of the $\order{B^3}$ terms in multifold semimetals.

In our calculations, we have assumed the same relaxation times to be applicable for all the bands.
In the future, we would like to improve our calculations by going beyond the relaxation-time approximation, which involves actually computing the collision integrals for all relevant scattering processes \cite{timm}, rather than just using phenomenological values of momentum-independent relaxation times.

Other directions worthwhile to be pursued are repeating our calculations for tilted RSW nodes \cite{timm, das-agarwal_omm, rahul-jpcm}, as tilting is expected in generic materials. Anisotropy arising from tilting of the dispersion \cite{emil_tilted, trescher17_tilted} is often neglected because it enters the Hamiltonian with an identity matrix, thus not affecting the eigenspinors
and, hence, the topology of the low-energy theory in the vicinity of the band-crossing point. For example, the BC or OMM of a Weyl or RSW cone does not depend on the tilt parameter. However, tilting does give rise to linear-in-$B$ terms, as seen in Refs.~\cite{rahul-jpcm, ips-tilted, das-agarwal_omm}. In fact, two of us have already computed \cite{ips-shreya} the signatures of topology in the terms linearly varying in $B$.
Furthermore, although we have considered the limit of weak non-quantizing magnetic field in this paper, we would like to study the influence of a strong quantizing magnetic field. This would involve incorporating the formation of the discrete Landau levels \cite{ips-kush, fu22_thermoelectric, staalhammar20_magneto, yadav23_magneto}. Lastly, if we want to move into the realm of linear and nonlinear response in the presence of strong disorder and/or strong interactions, we need to consider many-body techniques applicable for strongly-correlated systems \cite{ips-seb, ips_cpge, ips-biref, ips-klaus, rahul-sid, ipsita-rahul-qbt, ips-qbt-sc, ips-hermann-review}.

\appendix 

\section{Linear response from semiclassical Boltzmann equations}
 \label{secboltz}

In this appendix, we review the semiclassical Boltzmann formalism \cite{mermin, ips-kush-review, ips_rahul_ph_strain}, which is the used to determine the transport coefficients in the regime of linear response. There exists an externally-applied magnetic field $\mathbf B$, which we assume to be small in magnitude, leading to a small cyclotron frequency $\omega_c=e\,B/(m^*\, c) $ [where $m^* $ is the effective mass with the magnitude $\sim 0.11 \, m_e$ \cite{params2}, with $m_e$ denoting the electron mass]. This allows us to ignore quantized Landau levels, with the regime of validity of our approximations given by $\hbar \, \omega_c \ll \mu$, where $\mu$ is the Fermi level [i.e., the energy at which the chemical potential cuts the energy band(s)]. Furthermore, we will derive the expressions following from a relaxation-time approximation for the collision integral, which involves using a momentum-independent relaxation time. This implies that we will treat it as a phenomenological parameter. We assume that only intranode scatterings matter in the collision integral, such that we consider only the corresponding relaxation time $\tau$. In particular, we focus on the transport for a single node of chirality $\chi$. The derivation here closely follows the arguments outlined in Refs.~\cite{ips-kush-review, ips_rahul_ph_strain, ips-ruiz}

For a 3d system, we define the distribution function for the fermionic quasiparticles occupying a Bloch band labelled by the index $s$ at the node $\chi$, with the crystal momentum $\mathbf k$ and dispersion $\varepsilon_s (\mathbf k)$, by $ f_s^\chi ( \mathbf r , \mathbf k, t) $. Then
\begin{align}
dN_s^\chi = g_s \,f_s^\chi ( \mathbf r , \mathbf k, t) \,
\frac{ d^3 \mathbf k}{(2\, \pi)^3 } \,d^3 \mathbf r
\end{align}
is the number of particles occupying an infinitesimal phase space volume of $ dV_p = \frac{ d^3 \mathbf k}{(2\, \pi)^3 } 
\,d^3 \mathbf r $, centered at $\left \lbrace \mathbf r , \mathbf k \right \rbrace $ at time $t$. Here, $g_s$ denotes the degeneracy of the band.
In the presence of a nontrivial topology in the bandstructure, a nonzero orbital magnetic moment (OMM) is induced, and there appears a Zeeman-like correction to the energy due to the OMM, which we denote by $\eta_s^\chi (\mathbf k)$. Hence, we define the OMM-corrected dispersion and the corresponding modified Bloch velocity as
\begin{align} 
\label{eqtoten1}
	\xi^{\chi}_s ({\mathbf k})  = \varepsilon_{s}(\mathbf{k})  +  \eta^{\chi}_{s}(\mathbf{k}) 
\text{ and } \boldsymbol{w}^{\chi}_{s}(\mathbf{k})= 
 \nabla_{\mathbf{k}}   \varepsilon_{s}(\mathbf{k})  + \nabla_{\mathbf{k}}\eta^{\chi}_{s}(\mathbf{k})\,,
\end{align}
respectively. The Hamilton's equations of motion for the quasiparticles, under the influence of static electric ($\mathbf{E}$) and magnetic ($\mathbf{B}$) fields, are given by \cite{mermin, sundaram99_wavepacket, li2023_planar}
\begin{align}
\label{eqrkdot}
\dot {\mathbf r} &= \nabla_{\mathbf k} \, \xi^{\chi}_{s}  
- \dot{\mathbf k} \, \cross \, \mathbf \Omega^{\chi}_{s} 
 \text{ and }  
\dot{\mathbf k} = -\, e  \left( {\mathbf E}  + \dot{\mathbf r} \, \cross\, {\mathbf B} 
	\right ) \nn
\Rightarrow & \, \dot{\mathbf r}  = \mathcal{D}^{\chi}_{s} 
	\left[   \boldsymbol{w}^{\chi}_{s} + e \left ({\mathbf E}  \cross  
	 \mathbf \Omega^{\chi}_{s} \right )  + e   \left ( \mathbf \Omega^{\chi}_{s} \cdot 
	  \boldsymbol{w}^{\chi}_{s} \right  )  \mathbf B  \right] \text{ and }
\dot{\mathbf k}  = -\, e \,\mathcal{D}^{\chi}_{s} \,  \left[   {\mathbf E} 
+  \left (      \boldsymbol{w}^{\chi}_{s}   \cross  {\mathbf B} \right ) 
+  e  \left (  {\mathbf E}\cdot  {\mathbf B} \right )  \mathbf \Omega^{\chi}_{s}  \right].
\end{align}
where $-\, e$ is the charge carried by each quasiparticle.
Furthermore,
\begin{align}
\mathcal{D}^{\chi}_{s} =  \frac{1}
{ 1 + \, e  \left( \mathbf{B} \cdot\mathbf \Omega^{\chi}_{s}\right) }
\end{align} 
is the factor which modifies the phase volume element from $dV_p $ to $  (\mathcal{D}^{\chi}_{s})^{-1} \, dV_p$, such that the Liouville’s theorem (in the absence of collisions) continues to hold in the presence of a nonzero BC \cite{son13_chiral, xiao05_berry, duval06_Berry, son12_berry}.

Incorporating all these ingredients, the kinetic equation of the quasiparticles is finally given by \cite{lundgren14_thermoelectric, amit_magneto}
\begin{align}
\label{eqkin33}
& \mathcal{D}^{\chi}_{s} \,\left [ \partial_t  
+ \left \lbrace  \boldsymbol{w}^{\chi}_{s}
+ e 
\,{\mathbf E} \cross \mathbf{\Omega}^{\chi}_s
+ e \left(  {\mathbf \Omega}^{\chi}_{s} \cdot \boldsymbol{w}^{\chi}_{s}  \right)
	\mathbf{B} \right \rbrace 
	\cdot \nabla_{\mathbf r} 
-  e \left(  {\mathbf E} + \boldsymbol{w}^{\chi}_{s}  \cross {\mathbf B} 	\right) 
	\cdot \nabla_{\mathbf k} 
- e^2 \left ( {\mathbf E}  \cdot {\mathbf B} \right )
	\Omega^{\chi}_{s}   \cdot \nabla_{\mathbf k} \right ] f_s^\chi = I_{\rm coll}\,.
\end{align}
which results from the Liouville’s equation in the presence of scattering events. On the right-hand side, $I_{\rm coll}$
denotes the collision integral, which corrects the Liouville’s equation, taking into account the collisions of the quasiparticles. 

Let the contributions to the average DC electric and thermal current densities from the quasiparticles, associated with the band $s$ at the node with chirality $\chi$, be ${\mathbf J}_s^\chi$ and ${\mathbf J}^{{\rm th},\chi}_s$, respectively. The linear-response matrix, which relates the resulting generalized current densities to the driving electric potential gradient and temperature gradient, is expressed as
\begin{align}
\label{eqcur1}
\begin{bmatrix}
\left( J_s^\chi \right)_i \vspace{0.2 cm} \\
\left( {J}_s^{{\rm th},\chi}\right)_i 
\end{bmatrix} & = \sum \limits_j
\begin{bmatrix}
 \left( \sigma_{s}^\chi \right)_{ij} &  \left( \alpha_{s}^\chi \right)_{ij}
\vspace{0.2 cm}  \\
T  \left( \alpha_{s}^\chi \right)_{ij} &  \left( \ell_{s}^\chi \right)_{ij}
\end{bmatrix}
\begin{bmatrix}
E_j
\vspace{0.2 cm}  \\
- \, { \partial_{j} T } 
\end{bmatrix} ,
\end{align}
where $ \lbrace i, j \rbrace  \in \lbrace x,\, y, \, z \rbrace $ indicates the Cartesian components of the current density vectors and the response tensors in 3d.
Using the explicit forms of the solutions for $f_s^\chi$~\cite{nandy_2017_chiral, amit_magneto, nandy_thermal_hall, lundgren14_thermoelectric}, the electric and the thermal current densities are captured by the following:
\begin{align}
\label{eqcur}
{\mathbf J}^\chi_s
& =   -\, e \,  g_s  \int
\frac{ d^3 \mathbf k}{(2\, \pi)^3 } \,
(\mathcal{D}^{\chi}_{s})^{-1}    \, \dot{\mathbf r}
\,  f_s^\chi( \mathbf r , \mathbf k) \text{ and } 
\mathbf{J}^{\rm th,\chi}_s = g_s  \int
\frac{ d^3 \mathbf k}{(2\, \pi)^3 } 
\, (\mathcal{D}^{\chi}_{s})^{-1}   
\, \dot{\mathbf r} \,
\left( \xi^\chi_s - \mu \right)  f_s^\chi( \mathbf r , \mathbf k)\nn
\Rightarrow {\mathbf J}^\chi_s
& =   -\, e \,  g_s  \int
\frac{ d^3 \mathbf k}{(2\, \pi)^3 } \,
\left[   \boldsymbol{w}^{\chi}_{s} + e \, ({\mathbf E}  \cross  
	 \mathbf \Omega^{\chi}_{s})  + e   \, ( \mathbf \Omega^{\chi}_{s} \cdot 
	  \boldsymbol{w}^{\chi}_{s} ) \, \mathbf B  \right]
\,  f_s^\chi( \mathbf r , \mathbf k) \nn  \text{and } 
& \mathbf{J}^{\rm th,\chi}_s  = g_s  \int
\frac{ d^3 \mathbf k}{(2\, \pi)^3 } 
\left[   \boldsymbol{w}^{\chi}_{s} + e \, ({\mathbf E}  \cross  
	 \mathbf \Omega^{\chi}_{s}) \, + e   \, ( \mathbf \Omega^{\chi}_{s} \cdot 
	  \boldsymbol{w}^{\chi}_{s} ) \, \mathbf B  \right]
\left( \xi^\chi_s - \mu \right)  f_s^\chi( \mathbf r , \mathbf k)
\text{ [using Eq.~\eqref{eqrkdot}].}
\end{align}
Comparing with Eq.~\eqref{eqcur1}, we extract the final expressions for the linear-response coefficients. The notations $\sigma^\chi_s $ and $\alpha^\chi_s$ represents the magnetoelectric conductivity and the
magnetothermoelectric conductivity tensors, respectively. The latter determines the Peltier ($\Pi^\chi_s$), Seebeck ($ S^\chi_s $), and Nernst coefficients. The third tensor $\ell^\chi_s $ represents the linear response relating the thermal current density to the temperature gradient, at a vanishing electric field. $ S^\chi_s $ , $\Pi^\chi_s $, and the magnetothermal coefficient tensor $\kappa^\chi_s $ (which provides the coefficients between the heat current density and the temperature gradient at vanishing electric current) are related as \cite{mermin, ips-kush-review}:
\begin{align}
\label{eq:kappa}
\left( S_{s}^\chi \right)_ {ij} = \sum \limits_{ i^\prime}
\left( \sigma^\chi_s \right)^{-1}_{ i i^\prime }
\left ( \alpha^\chi_s\right)_{ i^\prime j} \, , \quad
 \left( \Pi_{s}^\chi \right)_ {ij} = T \sum \limits_{ i^\prime}
\left( \alpha^\chi_s \right)_{ i   i^\prime}   
\left(\sigma^\chi_s \right)^{-1}_{ i^\prime j} \,,\quad 
\left( \kappa_{s}^\chi \right)_ {ij} =
\left( \ell^\chi_s \right)_{ ij }
- T \sum \limits_{ i^\prime, \, j^\prime }
\left( \alpha^\chi_s \right)_{ i  i^\prime }
\left( \sigma_s^\chi \right)^{-1}_{  i^\prime  j^\prime }
\left( \alpha^\chi_s \right)_{ j^\prime j}  \,.
\end{align}
Since $\ell_s^\chi $ determines the first term in the magnetothermal coefficient tensor $\kappa_s^\chi $, here we will loosely refer to $\ell^\chi $ itself as the magnetothermal coefficient.

\subsection{Contributions from intranode scatterings}

We use the relaxation-time approximation, with only intranode and intraband scattering processes taken into account. The neglect of interband scatterings is justified if only pseudospin-conserving processes are allowed. Under these approximations/assumptions, the collision integral takes the form of
\begin{align}
I_{\rm coll} =   
	\frac{ f^{(0)}_{s, \chi} (\mathbf r,\mathbf k)- f_s^\chi(\mathbf r,\mathbf k, t) }
	{\tau } \,,
\end{align}
where the time-independent distribution function,
\begin{align}
\label{eqf0}
	f^{(0)}_{s, \chi} (\mathbf r,\mathbf k) \equiv 
	f_0 \big (\xi^\chi_s(\mathbf k) , \mu , T (\mathbf r) \big )
= \frac{1}
{ 1 + \exp [ \frac{ \xi^\chi_s(\mathbf k)-\mu} 
			{ T  (\mathbf r )}  ]}\,,
\end{align} 
describes a local equilibrium situation at the subsystem centred at position $\mathbf r$, at the local temperature $T(\mathbf r )$, and with a spatially uniform chemical potential $\mu $.

In order to obtain a solution to the full Boltzmann equation for small time-independent values of $\mathbf E$ and $\nabla_{\mathbf r} T$, we assume a small deviation, $\delta  f_s^\chi(\mathbf r,\mathbf k)$, from the equilibrium distribution of the quasiparticles. We have not included any explicit time-dependence in it since the applied fields and gradients are static. Hence, the nonequilibrium time-independent distribution function can be expressed as
\begin{align}
f_s^\chi(\mathbf r,\mathbf k, t) \equiv  f_s^\chi(\mathbf r,\mathbf k)
	=  f_0 +  \delta  f_s^\chi(\mathbf r,\mathbf k)\,,
\end{align} 
where we have suppressed showing explicitly the dependence of $f_0 $ on $\xi^\chi_s(\mathbf k)$, $\mu $, and $T (\mathbf r)$. At this point, the magnetic field is not assumed to be small, except for the fact that it should not be so large that the energy levels of the systems get modified by the formation of discrete Landau levels.

The gradients of the equilibrium distribution function $ f^{(0)}_{s, \chi}$ evaluate to
\begin{align}
\nabla_{\mathbf r}  f_0 
& =  \frac{ \xi_s^\chi - \mu } {T} 
\, \nabla_{\mathbf r} T 
\left( - \frac{\partial  f_0 } {\partial \xi_s^\chi } \right ) \text{ and }
\nabla_{\mathbf k}  f_0 
=   {\boldsymbol w}_s^\chi 
\, \frac{\partial  f_0 } {\partial \xi_s^\chi} \,.
\end{align}
We assume that $\delta f_s^\chi$ is of the same order of smallness as the external perturbations $\mathbf E $ and $\nabla_{\mathbf r} T$, and work in the linearized approximation (i.e., we keep terms upto the linear order in the ``smallness parameter''). Since the spatial gradient of $ f_0 $ is parallel to $\nabla_{\mathbf r} T$, and we limit ourselves to the situations where $\mathbf E $ and  $\nabla_{\mathbf r} T$ are applied along the same direction, the term $ e \left (  {\mathbf E} \cross \mathbf{\Omega}^{\chi }_s \right) \cdot \nabla_{\mathbf r} f_0 $ in Eq.~\eqref{eqkin33} vanishes. The term $ e \left (  {\mathbf E} \cross \mathbf{\Omega}^{\chi }_s \right) \cdot
\nabla_{\mathbf r} \delta f_s^\chi $ from Eq.~\eqref{eqkin33} also does not contribute, as it is of second order in smallness. Finally, we can write $ \delta  f_s^\chi (\mathbf r, \mathbf k)\simeq 
\delta  f_s^\chi ( \mathbf k) $ for spatially uniform $\mathbf E $ and $\nabla_{\mathbf r} T $. This leads to
the \textit{linearized Boltzmann equation}, given by
\begin{align}
\label{eqkin5}
& - D_{s}^\chi  \left [
\left \lbrace {\boldsymbol{w}}_s^\chi 
+e \left(
{\mathbf \Omega}^\chi_{s} \cdot {\boldsymbol{w}}_s^\chi   
 \right)  \mathbf B \right \rbrace
\cdot \left( \frac{  \xi_s^\chi - \mu } {T}  \, \nabla_{\mathbf r} T 
+ e \, \mathbf E
\right ) \right] 
	\frac{\partial  f_0 } {\partial \xi_s^\chi }
+  e \,  D_{s}^\chi \, {\mathbf B} \cdot
\left( {\boldsymbol{w}}_s^\chi   \cross \nabla_{\mathbf k} \right) 
\, \delta f_s^\chi (\mathbf k)
 = -\frac{\delta f_s^\chi (\mathbf k)} 
{\tau   } \,.
\end{align}
We want to solve the above equation for our planar Hall configurations by using an appropriate ansatz for $\delta f_s^\chi (\mathbf k)$.

\subsection{Solution for the Lorentz-force part} 
\label{appinter}

We now discuss how to include the Lorentz-force part. Here, we will set the $\nabla_{\mathbf r} T$ part to zero for the sake of brevity. The effect of a nonzero and uniform $\nabla_{\mathbf r} T$ can be easily inferred from the final solution for this case. To derive the coefficients of linear response, here we parametrize the nonequilibrium 
distribution function as \cite{deng2019_quantum}
\begin{align} 
\label{eqfpar}
f^{\chi}_{s} (\mathbf{k})   &=  f_{0} (\xi^{\chi}_{s}) + \delta f^{\chi}_{s} (\mathbf{k}) \,,
\quad
\delta f^{\chi}_{s} (\mathbf{k}) 
 = \left [- f_{0}^\prime (\xi^{\chi}_{s}) \right ] {\tilde g}^{\chi}_{s}  (\mathbf{k})\, ,
\end{align} 
where $\left [- f_{0}^\prime (\xi^{\chi}_s) \right ] {\tilde g}^{\chi}_s $ quantifies a small deviation of $f^{\chi}_{s} (\mathbf{k})$ from $f_0 (\xi^{\chi}_{s})$ due to the external probe fields, which are assumed to be spatially uniform and time-independent. As before, $f_0 (\xi^\chi_s )$ [cf. Eq.~\eqref{eqf0}] represents the equilibrium Fermi-Dirac distribution function for the quasiparticles occupying the $s^{\rm th}$ band at the node $\chi $. 
We define the Lorentz-force operator as 
\begin{align}
\hat{L} = (\boldsymbol{w}^{\chi}_{s} \cross \mathbf{B}) \cdot \nabla_{\mathbf{k}}\,.
\end{align} 
From Eq.~\eqref{eqkin5}, we get
\begin{align}
\label{eqkininter}
& e\,\mathcal{D}^{\chi}_{s} \left [
\left( \boldsymbol{w}^{\chi}_{s} 
+ \boldsymbol{W}^{\chi}_{s} \right ) \cdot \mathbf{E} 
\right ]
- e \,\mathcal{D}^{\chi}_{s} \, \hat{L} \,{\tilde g}^{\chi}_{s} (\mathbf{k})
=  \frac{1}{\tau } \left[
-\, {\tilde g}^{\chi}_{s} (\mathbf{k}) 
\right ] 
 \Rightarrow   
 \left (1 - e \, \tau \, \mathcal{D}^{\chi}_{s} \, \hat{L} \right )  {\tilde g}^{\chi}_{s}  
 = -  \,  e\,\tau \, \mathcal{D}^{\chi}_{s}  
\left[ \left ( \boldsymbol{w}^{\chi}_{s}
  + \boldsymbol{W}^{\chi}_{s} \right ) 
  \cdot \mathbf{E} \right ] ,
\end{align}
where
\begin{align}
\boldsymbol{w}^{\chi}_{s}(\mathbf{k})= 
	\boldsymbol{v}_s(\mathbf{k})  + \boldsymbol{u}^{\chi}_{s}(\mathbf{k})\,,\quad
 \mathbf{W}^{\chi}_{s} =  
 \mathbf{V}^{\chi}_{s} + \mathbf{U}^{\chi}_{s} \,,\quad
 \mathbf{V}^{\chi}_{s} = e \left ( \boldsymbol{v}_{s} 
 \cdot \boldsymbol{\Omega}^{\chi}_{s} \right ) \mathbf{B} 
 \,, \quad
 \mathbf{U}^{\chi}_{s} = e \left  ( \boldsymbol{u}^{\chi}_{s} \cdot
  \boldsymbol{\Omega}^{\chi}_{s} \right ) \mathbf{B} \,.
 \end{align}
This finally leads to
\begin{align}
\label{eqbolinter}
\frac{{\tilde g}^{\chi}_{s} (\mathbf{k}) } 
{ e \, \tau }
= -\, \sum_{n = 0}^{\infty}
\left (e \, \tau \, \mathcal{D}^{\chi}_{s} \right )^n \hat{L}^n 
\left [    \mathcal{D}^{\chi}_{s} \,
 \left ( \boldsymbol{w}^{\chi}_{s} 
+ \boldsymbol{W}^{\chi}_{s} \right ) \cdot \mathbf{E}   \right ],
\end{align}
which we solve for ${\tilde g}^{\chi}_{s} (\mathbf{k})$ recursively.

We can now expand the ${\tilde g}^{\chi}_{s} (\mathbf{k})$ upto any desired order in $ B$, in the limit of weak magnetic field, and obtain the current densities from Eq.~\eqref{eqcur}. In this paper, we are interested in terms upto cubic in $ B$. We observe that ${\tilde g}^{\chi}_{s} (\mathbf{k})$ includes the classical effect due to the Lorentz force.

\subsection{Expansion in $B$}

In order to obtain closed-form analytical expressions, we expand the $B $-dependent terms upto a given order in $B$, assuming it has a small magnitude, which is anyway required to justify neglecting the formation of the Landau levels.
With this in mind, we expand the Fermi-Dirac distribution as \cite{onofre}
\begin{align}
	f_0 (\xi^{\chi}_{s})  =  
 f_0  ( \varepsilon_s  )  + 
 \eta^{\chi}_{s} \, f^\prime_0  ( \varepsilon_s  ) + 
 \frac{(\eta^{\chi}_{s})^2}{2} \, f^{\prime \prime}_0  ( \varepsilon_s ) + 
  \frac{(\eta^{\chi}_{s})^3}{6} \, 
  f^{\prime \prime \prime}_0 ( \varepsilon_s ) + \mathcal{O} (B^4 ) \,.  
\end{align}
Analogously, $\mathcal{D}_s^\chi $ will be expanded as 
\begin{align}\mathcal{D}^{\chi}_{s} = 
\sum \limits_{n=0}^{\infty} 
\left [ -e \, \sum_j  (\Omega^{\chi}_{s})_j \, B_j  \right ]^n \,,
\end{align}
such that the final expressions are correct upto $\order{B^3}$.

Here, we show the explicit expression of Eq.~\eqref{eqsigLF} expanded upto order $B^3$.
Let us start with
\begin{align} 
\label{sigma_LF_appendix}
\left ( \sigma_s^{\chi} \right )_{i j} &=  - \,\epsilon_{j q r} \,
 \frac{e^3 \, \tau^2 \, 
s^3 \, v_0^3} {(2 \, \pi)^3}  
\sum_{a, {  \nu } , \zeta  }  \int d^3 \mathbf{k} \,
 \frac{\varrho_q \, \mathcal{I}_{ir}^{(a, {  \nu } , \zeta )}} 
 { \varepsilon_s ^5  }    
\; 
\text{ for } a \in \lbrace  v,\, u, \,V, \,U \rbrace
\text{ and } \lbrace {  \nu }, \,   \zeta  \rbrace 
\in \lbrace 0, \, \mathbb{Z}^+ \rbrace \,,\nn
\mathcal{I}_{ir}^{(a,{  \nu }, \zeta )} &= 
\frac{({\mathcal{D}^{\chi}_{s}} )^2}   { \zeta !}
\left [  \varepsilon_s^4 \, \mathcal{W}_{i}^{(a,{  \nu }+ \zeta )}  
 - 2 \, \varepsilon_s^2 \, \tilde{\mathcal{W}}_i^{(a,{  \nu }+ \zeta )} 
 + \check{\mathcal{W}}_i^{(a,{  \nu }+ \zeta )}  \right ] 
 \frac{\partial^{ \zeta +1}
  f_0 ( \varepsilon_s  ) }
 {\partial  \varepsilon_s^{ \zeta +1}} \,  B_r \,,\nn
\mathcal{W}_i^{(a,  \nu +1 )} & =
\eta^{\chi}_{s} \,\mathcal{W}_i^{(a,  \nu )}\, , \quad
 \tilde{\mathcal{W}}_i^{(a, \nu +1 )} = \lambda^{\chi}_{s}  \,\mathcal{W}_i^{(a, \nu )} 
 \,, \quad
\check{\mathcal{W}}_i^{(a, \nu + 2)} = 
(\lambda^{\chi}_{s})^2  \,\mathcal{W}_i^{(a, \nu)} \,. 
\end{align}
Explicitly, we have
\begin{align}
& \mathcal{W}_i^{(v, 0)} = (v^{\chi}_{s})_i \,, \quad  
\mathcal{W}_i^{(u, 1)} = (u^{\chi}_{s})_i \, , \quad  
\mathcal{W}_i^{(V, 1)} = (V^{\chi}_{s})_i \,, \quad
 \mathcal{W}_i^{(U, 2)} = (U^{\chi}_{s})_i \,, \quad
\mathcal{W}_i^{(u, 0)} =  \mathcal{W}_i^{(V, 0)}
 = \mathcal{W}_i^{(U, 0)} =  \mathcal{W}_i^{(U, 1)} = 0\, ,
\end{align}
needed for the final forms upto $\order{B^3}$.
Here, $ \zeta +  \nu $ denotes that $ B^{\zeta +  \nu} $ appears in that term. Keeping terms upto $ \zeta =1$ gives us the expressions upto order $ B^3 $. We also need to use the series expansion
\begin{align}
({\mathcal{D}^{\chi}_{s}} )^2 = 
\sum_{ n  = 1}^{\infty} n  \left [- \, e 
\sum_{i \in \lbrace x,y, z\rbrace } 
(\Omega^{\chi}_{s})_i \, B_i  \right ]^{n-1} \,.
\end{align}
Putting all the pieces together, we finally get
\begin{align} 
& \sum_{a,\nu ,\zeta} \mathcal{I}_{ir}^{(a,\nu ,\zeta )}   =
t^{(1)}_{ir} + t^{(2)}_{ir} + t^{(3)}_{ir} + \order{B^4} \,,\nn
& t^{(1)}_{ir}  =  \varepsilon_s^4 \, \mathcal{W}_{i}^{(v,0)}  \,  
f^\prime_0 \left ( \varepsilon_s \right )   B_r \,, \nn
& t^{(2)}_{ir}  = -\, 2 \, e \sum_{i'} 
(\Omega^{\chi}_{s})_{i'} \, B_{i'} \, t_{ir}^{(1)}  
+ \Bigg [  
\left \lbrace \varepsilon_s^2 \,
\left (  \mathcal{W}_i^{(u,1)} 
+   \mathcal{W}_i^{(V,1)} \right )
- 2 \, \vartheta \sum_{i'} \varrho_{i'} {B}_{i'} \,  \mathcal{W}_i^{(v,0)} \right \rbrace
 \varepsilon_s^2 \, f^\prime_0 (\varepsilon_s) 
\nn &
\hspace{5 cm } - \sum_{j'} (m^{\chi}_{ s })_{j'} \, B_{j'} \, \varepsilon_s^4 \, \mathcal{W}_i^{(v,0)} \, 
f^{\prime \prime}_0 (\varepsilon_s)   \Bigg ] B_r  \,,\nn
& t^{(3)}_{ir} =  - \,e^2 \sum_{ i', \, j'} (\Omega^{\chi}_{s})_{i'} 
 \, (\Omega^{\chi}_{s})_{j'} \, B_{i'} 
 \, B_{j'} \,  t^{(1)}_{ir} 
 -  2 \, e  \sum_{i'} (\Omega^{\chi}_{s})_{i'} \, B_{i'} \, t^{(2)}_{ir}   
\nn & \hspace{ 1 cm }
+  \Bigg [ \Bigg \{  \varepsilon_s^4 \, \mathcal{W}_i^{(U,2)}
 - 2 \,\vartheta \sum_{i'} \varrho_{i'} {B}_{i'} \, \varepsilon_s^2 \left ( \mathcal{W}_i^{(u,1)} 
 +   \mathcal{W}_i^{(V,1)}  \right ) 
+ \vartheta^2 \sum_{i' ,\,j'} \varrho_{i'} {B}_{i'} \varrho_{j' } \, B_{j'}  
\mathcal{W}_i^{(v,0)}  \Bigg \}  f^\prime_0 (\varepsilon_s)  
 \nn & \hspace{ 1.7 cm }
  - \sum_{j'} (m^{\chi}_{ s })_{j'} \, B_{j'}  \left \lbrace  \, \varepsilon_s^2 
  \left ( \mathcal{W}_i^{(u,1)}   
 + \mathcal{W}_i^{(V,1)}  \right )
  - 2 \, \vartheta \sum_{i'}  
  \varrho_{i'} {B}_{i'} \, \mathcal{W}_i^{(v, 0 )} \right \rbrace 
  \varepsilon_s^2 \, 
 f^{\prime \prime}_0 ( \varepsilon_s)  
 \nn  &  
\hspace{ 1.75 cm} 
 + \left \lbrace \frac{\varepsilon_s^4}{2} \, \mathcal{W}_i^{(v,0)} 
 \sum_{i', \,j'} (m^{\chi}_{ s })_{i'} \, B_{i'} \, (m^{\chi}_{ s })_{j'} \, B_{j'} 
 \right \rbrace f^{\prime \prime \prime}_0 ( \varepsilon_s) 
 \Bigg ] B_r \,,
\end{align}
where $ \vartheta = 2 \, \chi \, e\, s \,  {\mathcal{G}}_s \, v_0^2$.
Plugging this in into the integrand of Eq.~(\ref{sigma_LF_appendix}), it can be expanded in small $1/(\beta \, \mu)$, using the Sommerfeld expansion [cf. Appendix~\ref{appint}], to get the final expression.

\section{Sommerfeld expansion} 
\label{appint}

Throughout this paper, we have to deal with integrals of the form:
\begin{align}
I = \int \frac{d^{3} {\bf{k}} }{(2 \pi )^{3}} \, F ({\bf{k}} , \xi_s^{\chi} ) \,
 f_{0}^\prime (\xi_s^{\chi})  \,,
\end{align}
where $\xi_s^{\chi} =  \varepsilon_{s}(\mathbf{k}) + \eta^{\chi}_{s}(\mathbf{k}) $. 
We focus on the conduction bands, such that only the positive values of $s$ are relevant, which we denote by $\tilde s$.
Exploiting the spherical symmetry of the system, we introduce the spherical polar coordinates such that
\begin{align}
\label{eqpolar} 
k_{x} = \frac{\epsilon \cos \phi \sin \gamma} 
 { \tilde s\, v_0}\,, \quad
 k_{y} = 
\frac{\epsilon \sin \phi \sin \gamma} { \tilde s\, v_0}\,, 
\quad k_{z} 
 = \frac{\epsilon \cos  \gamma} { \tilde s\, v_0} \,.
\end{align}
The limits are: $\epsilon \in [0, \infty )$, $\phi \in [0, 2 \pi )$, and $\gamma \in [0, \pi ]$. The Jacobian of the transformation is $\mathcal{J} (\epsilon , \gamma ) =   \frac{\epsilon^2 \, \sin \gamma}
{ {\tilde s}^3 \, v_0^3} $. This leads to 
\begin{align}
\int_{- \infty}^{ \infty} d \mathbf{k }  \rightarrow   \int_{- \infty}^{ \infty} d \epsilon  
\int_{0}^{ 2 \pi} d \phi \,   \int_{0}^{  \pi} d \gamma \,   \mathcal{J} (\epsilon , \gamma ) 
\text{ and } 
\xi^{\chi}_{\tilde s}(\mathbf{k})   \rightarrow  
\xi^{\chi}_{\tilde s}(\mathbf{\epsilon})   
=   \epsilon + \eta^{\chi}_{\tilde s}(\epsilon) \,.
\end{align}
With the implementation of the above coordinate transformation, we have
\begin{align}
I & = \frac{1}{(2 \pi )^{3}}  \int_{- \infty}^{ \infty} d \epsilon  
\int_{0}^{ 2 \pi} d \phi \,   \int_{0}^{  \pi} d \gamma \, 
\mathcal{F} ( \epsilon, \phi , \gamma , \xi_{\tilde s}^{\chi} )   \, 
 f^\prime_{0} (\xi_{\tilde s}^{\chi}) 
\quad \text{ [where } \mathcal{F} ( \epsilon, \phi , \gamma , \xi_{\tilde s}^{\chi} )
=  \mathcal{J} (\epsilon , \gamma ) \, 
F ( \epsilon, \phi , \gamma , \xi_{\tilde s}^{\chi} ) ]
\nn & = \frac{1}{(2 \pi )^{3}}  \int_{0}^{\infty} d \epsilon \,   
\mathcal{K} (\chi, \epsilon )   \,  f^\prime_{0} (\xi_{\tilde s}^{\chi}) 
\quad \text{ [where }
\mathcal{K} ( \chi, \epsilon  )
 = \int_{0}^{2 \, \pi} d \phi \,  \int_{0}^{\pi} d \gamma \, 
 \mathcal{F} ( \epsilon, \phi , \gamma , \xi_{\tilde s}^{\chi} ) ]\, .
\end{align} 
This remaining part can be calculated using the Sommerfeld expansion \cite{mermin} under the condition $ 1/(\beta \, \mu) \ll 1$. The integral will turn out to consist of terms of the form
\begin{align}
\int_{0}^{ \infty} d \epsilon   \,   
 \epsilon^{n}   \left [ - f^\prime_{0} ( \epsilon )\right ] 
=  \Upsilon_{n} (\mu ) \text{ for } n \in \mathbb{Z} \,,
\end{align}
which, upon using the Sommerfeld expansion, yields
\begin{align}
\Upsilon_{n} (\mu )=  \mu^{n} \,  \left[ 1 + \frac{\pi^2 \,  n \,  (n-1)}
{6 \left(  \beta \,  \mu \right)^{2} } 
+ \order{\left( {\beta \,\mu}\right)^{-3}} \right] .
\end{align}

For higher-order derivatives we have 
\begin{align}
\int_{0}^{ \infty} d \epsilon    \,   \epsilon ^{n}  \,(-1)^{{ \lambda }+1}  \,
\frac{\partial^{{ \lambda }+1} 
	\, f_{0} ( \epsilon  ) } { \partial \epsilon ^{{ \lambda }+1} }  
= \frac{n!}{(n-{ \lambda })!}  \, \Upsilon_{n-{ \lambda }} (\mu )\,. 
\label{Identity3}
\end{align}

For the thermoelectric and thermal tensors, we need to use the identity 
\begin{align}
\int_{0}^{ \infty} d \epsilon   \, \epsilon^{n}  \,  (\epsilon - \mu ) 
\,  (-1)^{{ \lambda }+1}  
\, \frac{\partial^{{ \lambda }+1} f_{0} ( \epsilon ) }
{ \partial \epsilon^{{ \lambda }+1} } 
= \frac{(n+1)!}{(n+1-{ \lambda })!}  \, 
\Upsilon_{n+1-{ \lambda }} (\mu )- \mu \, \frac{n!}{(n-{ \lambda })!}  
\, \Upsilon_{n-{ \lambda }} (\mu ) \,. 
\label{Identity4}
\end{align}

\section{Magnetoelectric conductivity} 
\label{appLMC}

In this appendix, we outline the details of the steps to obtain the various parts of the magnetoelectric conductivity
tensor. This is determined by the electric current density expression shown in Eq.~\eqref{eqcur} [after setting $g_s =1 $], i.e., 
\begin{align}
\label{eqcurcond}
 {\mathbf J}^\chi_s
& =   -\, e   \int
\frac{ d^3 \mathbf k}{(2\, \pi)^3 } \,
\left[   
e \, ({\mathbf E}  \cross   \mathbf \Omega^{\chi}_{s})  
+ \boldsymbol{w}^{\chi}_{s} 
+ e   \, ( \mathbf \Omega^{\chi}_{s} \cdot 
	  \boldsymbol{w}^{\chi}_{s} ) \, \mathbf B  \right]
\,  f_s^\chi( \mathbf r , \mathbf k) \,.
\end{align}

\subsection{Intrinsic anomalous-Hall part}   
\label{appsigmaAH}

From the term proportional to $({\mathbf E}  \cross   \mathbf \Omega^{\chi}_{s})$ in the integrand of Eq.~\eqref{eqcurcond},
we get the linear-response current density as
\begin{align}
 {\mathbf J}^{\chi, {\text{AH}}}_s
& =   -\, e^2   \int
\frac{ d^3 \mathbf k}{(2\, \pi)^3 } \,
\left[   ({\mathbf E}  \cross   \mathbf \Omega^{\chi}_{s})   \right]
\,  f_0( \xi^\chi_s ) \,,
\end{align}
which gives the intrinsic anomalous-Hall term. The corresponding components of the conductivity are given by [cf. Eqs.~\eqref{eqsigmatot}]
\begin{align}
(\sigma^{\chi, {\text{AH}}}_s)_{ij} & = - \, e^2 \,\epsilon_{ijl}
\int \frac{  d^3 {\mathbf k} }  {(2 \,\pi )^{3} } \, 
(\Omega^{\chi}_s )^l  
\left [ f_0  (\varepsilon_s )  
+\eta^{\chi}_{s}  \, f_0^\prime (\varepsilon_s ) + \frac{1}{2} \, 
\left ({\eta^{\chi}_s}\right )^2 \,
f_0^{ \prime \prime} (\varepsilon_s ) +  \frac{1}{6} \, 
\left ({\eta^{\chi}_s}\right )^3 \,
f_0^{ \prime \prime \prime} (\varepsilon_s ) + \order{ B^4}  \right ]\,,
\end{align}
whose diagonal components (i.e., the $ii$-components) are automatically zero because of the Levi-Civita function. A nonzero OMM generates $B$-dependent terms. The first and the third terms will always vanish (for both the in-plane and out-of-plane transverse components) because of the vanishing of the integrals (the integrand being odd in $\mathbf k $). For our configuration with $\mathbf E$ and $\mathbf B$ confined to the $xy$-plane, we have
\begin{align}
(\sigma^{\chi, {\text{AH}}}_s)_{yx} = - \frac{e^2 \, \chi } {8\, \pi^3 \, v_0}  \, 
\int_{- \infty}^{ \infty} d \epsilon  
\int_{0}^{ 2 \pi} d \phi  \int_{0}^{  \pi} d \gamma     
  \sin{\gamma} \cos{\gamma}  
 \left [ \eta^{\chi}_s  \, f_0^\prime (\varepsilon_s )  +  \frac{1}{6} \, 
\left (\eta^{\chi}_s\right )^3 \,
f_0^{ \prime \prime \prime} (\varepsilon_s )  \right] 
 = 0
  =  (\sigma^{\chi, \text{AH}}_s)_{xy} \,.
\end{align}
Only the following out-of-plane component is nonzero: 
\begin{align}
(\sigma^{\chi, {\text{AH}}}_s)_{zx} & =  \frac{e^2 \, \chi }
{8 \, \pi^3 \, v_0}  \, 
\int_{- \infty}^{ \infty} d \epsilon  
\int_{0}^{ 2 \pi} d \phi   \int_{0}^{  \pi} d \gamma  
 \sin^2 \gamma   \sin{\phi} 
  \left [ \eta^{\chi}_s  \, f_0^\prime (\varepsilon_s )  
+  \frac{1}{6} \, 
\left (\eta^{\chi}_s\right )^3 \,
f_0^{ \prime \prime \prime} (\varepsilon_s )  \right ]  .
\end{align}  
This leads to the final expression shown in Eq.~\eqref{eqah}.

\subsection{Non-anomalous-Hall contribution with intranode-only scatterings}   
\label{appbcomm}

The non-anomalous-Hall contribution (not including the Lorentz-force contribution) with intranode-only scatterings is obtained by picking up the $n=0$ term on the right-hand side of Eq.~\eqref{eqbolinter}, i.e., by using 
\begin{align}
 \delta f^{\chi}_{s} (\mathbf{k})
=  e \, \tau \, \mathcal{D}^{\chi}_{s}\left [     \left ( \boldsymbol{w}^{\chi}_{s} 
+ \boldsymbol{W}^{\chi}_{s} \right ) \cdot \mathbf{E}   \right ] f_0^\prime (\xi^\chi_s) \,.
\end{align}
We plug this in into the non-anomalous-Hall part of Eq.~\eqref{eqcurcond} to obtain 
\begin{align}
 {\mathbf{\bar J}}^\chi_s
& =   -\, e^2\,\tau   \int
\frac{ d^3 \mathbf k}{(2\, \pi)^3 } \,
\left[    \boldsymbol{w}^{\chi}_{s} 
+ e   \, ( \mathbf \Omega^{\chi}_{s} \cdot 
	  \boldsymbol{w}^{\chi}_{s} ) \, \mathbf B  \right]
 \mathcal{D}^{\chi}_{s}\left [     \left ( \boldsymbol{w}^{\chi}_{s} 
+ \boldsymbol{W}^{\chi}_{s} \right ) \cdot \mathbf{E}   \right ] f_0^\prime (\xi^\chi_s) \,,
\end{align}
leading to
\begin{align}
\left(\bar \sigma^{\chi}_{s} \right)_{i j} 
= - \,e^2 \, \tau  
\int \frac{ d^3 \mathbf k}{(2\, \pi)^3 } \, \mathcal{D}^{\chi}_{s} 
\left[  (w^{\chi}_{s})_i \, + (W^{\chi}_{s})_i \right ]
\left [ (w^{\chi}_{s})_j \, + (W^{\chi}_{s})_j \right] \, f^\prime_0 (\xi^{\chi}_{s})  \,.
\end{align}
This is the expression shown in Eq.~\eqref{eqsigmatot} of the main text.

We want to compute here the ${\bar \sigma}^{\chi}_{s}  $-part, after dividing it up as
\begin{align}
{\bar \sigma}^{\chi}_{s}  =  
\sigma^{\chi, \text{Drude}}_{s} + \sigma^{\chi, \text{BC}}_{s} 
+ \sigma^{\chi, m}_{s} \,,
\end{align} 
where (1) the first part is the one which is independent of $\mathbf B $, also known as the Drude contribution;
(2) the second part arises solely due to the effect of the BC and survives when OMM is set to zero; and
(3) the third part is the one which goes to zero if OMM is ignored.

\subsubsection{Drude part}

Explicity, the Drude part is expressed as
\begin{align}
\left ( \sigma^{\chi, \text{Drude}}_{s} \right )_{i j} = 
- \, \frac{e^2 \, \tau }{(2 \, \pi)^3} 
\int d^3 \mathbf{k} \, \mathcal{L}_{i j}^{(0)} \,  
f^\prime_0  (\varepsilon_{s} ) \,,
\quad \mathcal{L}_{i j}^{(0)} =  (v^{\chi}_{ s })_{i} \, (v^{\chi}_{ s })_{j}\,.
\end{align}
The isotropy of the RSW bands, in the vicinity of a node, ensures that the off-diagonal terms vanish, i.e., $\left ( \sigma^{\chi, \text{Drude}}_{s} \right )_{i j} \propto \delta_{ij} $. This leaves only the longitudinal components of the tensor, which are given by
\begin{align}
(\sigma^{\chi, \text{Drude}}_{s})_{i i} = - \frac{e^2 \, \tau }{(2 \, \pi)^3} 
\int d^3 \mathbf{k} \, [ (v^{\chi}_{ s })_{i} ]^2  \,  f^\prime_0  (\varepsilon_{s} ) . 
\end{align} 

\subsubsection{BC-only part (no OMM)}

The BC-only part is given by 
\begin{align}
(\sigma^{\chi, \text{BC}}_{s})_{i j} &= 
- \, \frac{e^2 \, \tau }{(2 \, \pi)^3} 
\int d^3 \mathbf{k} \,  \mathcal{M}_{i j}   \,  f^\prime_0  (\varepsilon_{s} )\,,
\quad
\mathcal{M}_{i j} = \left[  \mathcal{D}^{\chi}_{s} 
\left \lbrace   \frac{  \mathcal{L}_{i j}^{(0)} } {2}
 + (v^{\chi}_{ s })_{i} \, (V^{\chi}_{ s })_{j} 
 + \frac{1}{2} \, (V^{\chi}_{ s })_{i} \, (V^{\chi}_{ s })_{j} 
\right \rbrace  + i \leftrightarrow j \right ] -\mathcal{L}_{i j}^{(0)}  , 
\end{align}
which is symmetric in the indices $i$ and $j$.
Here, we find that
\begin{align}
\label{eqmij}
 \mathcal{M}_{i j}  
&  =  \Big [
\frac{e \,  \mathcal{L}_{i j}^{(0)} } {2} 
 \Big \lbrace  -\sum_q (\Omega^{\chi}_{s})_q B_{q}  
 + e  \,  \sum_{q, r} ( \Omega^{\chi}_{s})_q \, 
  ( \Omega^{\chi}_{s})_r \, B_{q} \, B_{r} \Big \rbrace  
+ e \,  (v^{\chi}_{ s })_{i}  
 \Big \lbrace  \sum_{q} (v^{\chi}_{ s })_{q} \,  
(\Omega^{\chi}_{ s })_{q} \, B_j \nonumber \\
& \qquad - e  \sum_{q, r} (v^{\chi}_{ s })_{q} \,  (\Omega^{\chi}_{ s })_{q} \,  
 (\Omega^{\chi}_{ s })_{r} \, B_{r} \, B_j \Big \rbrace  
+ \frac{e^2}{2}   \sum_{q, r} (v^{\chi}_{ s })_{q} \, (v^{\chi}_{ s })_{r} \, 
 (\Omega^{\chi}_{ s })_{q} \,   (\Omega^{\chi}_{ s })_{r} \, B_i \, B_j  
 \Big ] + i \leftrightarrow j 
\nn & = \mathcal{M}_{i j}^{(1)}  + \mathcal{M}_{i j}^{(2)} +\order{B^3} , \nn
\mathcal{M}_{i j}^{(1)} & = \Big[- \frac{e}{2}  \,
(v^{\chi}_{ s })_{i} \, (v^{\chi}_{ s })_{j}  
\sum_q (\Omega^{\chi}_{s})_q \,B_{q} 
+ e \,  (v^{\chi}_{ s })_{i}   
  \sum_{q} (v^{\chi}_{ s })_{q} \,  (\Omega^{\chi}_{ s })_{q} \, B_j \Big] 
+ i \leftrightarrow j  \,, \nn
\mathcal{M}_{i j}^{(2)} & =
\Big [ \frac{e^2}{2} \, (v^{\chi}_{ s })_{i} \, 
(v^{\chi}_{ s })_{j} 
   \sum_{q, r} ( \Omega^{\chi}_{s})_q \,  ( \Omega^{\chi}_{s})_r \, B_{q} \, B_{r} 
- e^2 \,  (v^{\chi}_{ s })_{i}   
 \sum_{q, r} (v^{\chi}_{ s })_{q} \,  (\Omega^{\chi}_{ s })_{q} \,   
(\Omega^{\chi}_{ s })_{r} \, B_{r} \, B_j  \nonumber  \\
& \qquad + \frac{e^2}{2}   \sum_{q, r} (v^{\chi}_{ s })_{q} \, (v^{\chi}_{ s })_{r} \,  (\Omega^{\chi}_{ s })_{q} \,   (\Omega^{\chi}_{ s })_{r} \, B_i \, B_j  \Big ] + i \leftrightarrow j \,.
\end{align}
The rotational symmetry of the system makes the part with $\mathcal{M}_{i j}^{(1)}$ vanish. In fact, in general, $(\sigma^{\chi, \text{BC}}_{s})_{i j}$ consists of terms which contain only even powers of $B$. Therefore, the $\order {B^3}$ term also vanishes, which implies that the above expression is correct upto $\order {B^3}$.

\subsubsection{Part with the integrand proportional to nonzero powers of OMM}

The OMM shifts the dispersion by $\eta^{\chi}_{s} = - \sum_{i} (m^{\chi}_{s})_i  \, B_i  $ [cf. Eq.~\eqref{eqomme}]. Let us define 
\begin{align}
(u^{\chi}_{s})_{i} &= \sum_j (\mathcal{U}^{\chi}_{s})_{i j} \, B_{j} \,,
\end{align}
where
\begin{align}
\mathcal{U}^{\chi}_{s} =
\begin{bmatrix}
	\Delta_{11} &   \Delta_{12} &  \Delta_{13}  \\
	\Delta_{21} &   \Delta_{22} &   \Delta_{23}  \\
	\Delta_{31} &   \Delta_{32} &  \Delta_{33}
\end{bmatrix} ,
\end{align}
\begin{align}
\Delta_{11}(\epsilon, \gamma, \phi) &= - \,
\chi \, e \, \mathcal{G}_s \, s^2 \, v_0^3 \;
 \frac{  \sin^2 \gamma  \cos (2 \phi )
 -\cos^2 \gamma  } {\epsilon^2}\,,\quad
\Delta_{22} (\epsilon, \gamma, \phi) = 
 \Delta_{11}(\epsilon, \gamma, \frac{\pi}{2} - \phi) \,,\nonumber \\
\Delta_{33} (\epsilon, \gamma, \phi) &= -\,
 \Delta_{11}(\epsilon, \gamma, 0)  \,,\quad
\Delta_{12} (\epsilon, \gamma, \phi) = \Delta_{21} (\epsilon, \gamma, \phi) 
= - \chi \, e \, \mathcal{G}_s \, s^2 \, v_0^3 \;   
\frac{\sin^2 \gamma  \,\sin (2 \phi )} {\epsilon^2} \,,\nonumber \\
\Delta_{13} (\epsilon, \gamma, \phi) &= \Delta_{31} (\epsilon, \gamma, \phi) 
= \frac{\Delta_{12} }{\tan{\gamma} \sin{\phi}} \,,\quad
\Delta_{23} (\epsilon, \gamma, \phi) = \Delta_{32} (\epsilon, \gamma, \phi)
 = \frac{\Delta_{12} }{\tan{\gamma} \cos{\phi}} \,.
\end{align} 
Here, we have used the spherical polar coordinates, defined in Eq.~\eqref{eqpolar}, to write the matrix elements of $\mathcal{U}^{\chi}_{s}$ in a compact form.

We can now express the relevant part of the conductivity as 
\begin{align}
& (\sigma^{\chi, m}_{s})_{i j}  = -  e^2 \, \tau 
\int \frac{d^3 \mathbf{k}}{(2 \, \pi)^3}   \left[ \mathcal{S}_{i j}  \, f^{\prime}_0 (\varepsilon_{s} )  +  \left(  \mathcal{P}_{i j} + \mathcal{Q}_{i j}  +\mathcal{R}_{i j} \right)   \, f^{\prime \prime}_0 (\varepsilon_{s} ) 
+ \frac{1}{2} \, \mathcal{T}_{i j}  \, f^{\prime \prime \prime}_0 (\varepsilon_{s} ) 
\right] \,,
\end{align}
where
\begin{align} 
\label{eqpqrs}
& \mathcal{P}_{i j} = \left [ - \frac{1}{2} 
\sum_{ i'} (v^{\chi}_{ s })_{i} \, (v^{\chi}_{ s })_{j} \,
 (m^{\chi}_{ s })_{i'} \, B_{i'} \right ] + i \leftrightarrow j \,,\nn
& \mathcal{Q}_{i j} =  
\left [ \frac{e}{2} 
\, (v^{\chi}_{ s })_{i} \, (v^{\chi}_{ s })_{j}  
  \sum_{ i',\, j'} (\Omega^{\chi}_{s})_{i'} \,  (m^{\chi}_{s})_{j'} \, B_{i'} \, B_{j'}  
- e \,  (v^{\chi}_{ s })_{i}  
 \sum_{i' ,\,j'} (v^{\chi}_{ s })_{i'} 
\,  (\Omega^{\chi}_{ s })_{i'} \, (m^{\chi}_{s})_{j'} \, B_{j'} \, B_j   \right ] 
+ i \leftrightarrow j \,,\nn
& \mathcal{R}_{i j} =  \left [ - \, (v^{\chi}_{ s })_{i} 
 \sum_{ i',\, j'} (\mathcal{U}^{\chi}_{s})_{j {i'}}  
 \, (m^{\chi}_{s})_{j'} \, B_{i'} \, B_{j'}  
   \right ] + i \leftrightarrow j\,, \nn
 &  \mathcal{S}_{i j}  =   \Bigg[
  (v^{\chi}_{ s })_{i} 
  \sum_{i'} (\mathcal{U}^{\chi}_{s})_{j i'} \, B_{i'} 
 - e \,(v^{\chi}_{ s })_{i} 
 \sum_{i'} (\mathcal{U}^{\chi}_{s})_{j i'} \, B_{i'}   
 \sum_{j'} (\Omega^{\chi}_{s})_{j'} B_{j'}   
 +   \frac{1}{2} \, \sum_{i', \, j'} (\mathcal{U}^{\chi}_{s})_{i i'} 
 \,  (\mathcal{U}^{\chi}_{s})_{j j'} \, B_{i'} \, B_{j'} \nonumber \\
 & \hspace{1.3 cm }
 + e \, (v^{\chi}_{ s })_{i} 
  \sum_{i', \, j'} (\mathcal{U}^{\chi}_{s})_{i' j'} \, 
  (\Omega^{\chi}_{s})_{i'} \, B_j \,  B_{j'} 
  + e  \sum_{i' ,\,j'} (\mathcal{U}^{\chi}_{s})_{i i'}  \, (v^{\chi}_{s})_{j'} \, 
  (\Omega^{\chi}_{s})_{j'} \, B_{j} \, B_{i'}    + i \leftrightarrow j  \Bigg]   + i \leftrightarrow j
\nn  &  \mathcal{T}_{i j} =  \left [ \frac{1}{2} \, 
\sum_{ i',\, j'} (v^{\chi}_{ s })_{i} \, (v^{\chi}_{ s })_{j} \, (m^{\chi}_{s})_{i'} 
\, (m^{\chi}_{s})_{j'} \, B_{i'} \, B_{j'}  
\right ] + i \leftrightarrow j  \,,
\end{align}
on keeping terms upto quadratic order in $B$. The isotropy of the system makes the parts with odd powers of $B$ vanish. As a result, in general, terms with only even powers of $B$ survive. Therefore, the $\order {B^3}$ term also vanishes, which implies that the above expression is correct upto $\order {B^3}$.

\subsection{Lorentz-force contribution}
\label{applor}

The leading-order contribution from the Lorentz-force part is obtained by picking up the
$n=1$ term on the right-hand side of Eq.~\eqref{eqbolinter}, i.e., by using 
\begin{align}
 \delta f^{\chi}_{s} (\mathbf{k})
=  e^2 \, \tau^2  \, \mathcal{D}^{\chi}_{s} 
\, f_0^\prime (\xi^\chi_s) \,\hat{L}
\left [  \mathcal{D}^{\chi}_{s}   \left ( \boldsymbol{w}^{\chi}_{s} 
+ \boldsymbol{W}^{\chi}_{s} \right ) \cdot \mathbf{E}   \right ] .
\end{align}
Plugging it in into the non-anomalous-Hall part of Eq.~\eqref{eqcurcond}, we obtain the current density
\begin{align} 
\label{eqcurlf1}
{\mathbf J}^{\chi, \rm LF}_{s} = 
-\,e^3 \,  \tau^2 \int \frac{d^3 \mathbf{k}} {(2 \, \pi)^3} 
\, {\mathcal{D}^{\chi}_{s}}  \, f_{0}^\prime (\xi^{\chi}_{s})
\left [ \boldsymbol{w}^{\chi}_{s} 
+ \boldsymbol{W}^{\chi}_{s}
\right ] 
\hat{L} 
\left [  \mathcal{D}^{\chi}_{s} \, 
\left \lbrace \left ( \boldsymbol{w}^{\chi}_{s} 
+ \boldsymbol{W}^{\chi}_{s} \right ) \cdot \mathbf{E} \right \rbrace 
 \right ]  .
\end{align}
This is the classical Hall current density due to the Lorentz force.

Using various vector identities, we get
\begin{align}
& \hat{L}  \left ( \boldsymbol{W}^{\chi}_{s}  \cdot 
\mathbf{E} \right )  = 0 \,, 
\quad \hat{L} \,  {\mathcal{D}^{\chi}_{s}} = 0 \,,
\quad \hat{L} \left ( \boldsymbol{w}^{\chi}_{s} \cdot \mathbf{E} \right )
 = \frac{ s^3 \, v_0^3 } {  \varepsilon_s^5} 
 \left(  \varepsilon_s^2 - \lambda_s^{\chi} \right)^2 
\left  ( \boldsymbol \varrho \cross \mathbf{B} \right )  \cdot \mathbf{E} \,,
\end{align}
where
\begin{align}
 \lambda_s^{\chi} = \vartheta \sum_q \varrho_q \, {B}_q\,,\quad
 \boldsymbol{\varrho} = \cos{\phi} \sin{\gamma}\, \boldsymbol {\hat  x}
 + \sin{\phi} \sin{\gamma}\, \boldsymbol {\hat  y}
 + \cos{\gamma} \boldsymbol {\hat  z} \,, \quad
 \vartheta = 2 \, \chi \, e\, s \,  {\mathcal{G}}_s \, v_0^2 \,.
\end{align}
Furthermore, $\phi $ and $\gamma $ refer to the azimuthal and polar angles of the spherical polar coordinates, which the components of $\mathbf k$ are transformed to, as shown in Appendix~\ref{appint}.
This leads to the simplification of Eq.~\eqref{eqcurlf1} into
\begin{gather}
{\mathbf J}_s^{\chi, \rm LF} =
 - \, e^3 \, \tau^2 \, s^3 \,v_0^3
  \int \frac{d^3 \mathbf{k} } {(2 \, \pi)^3} \, 
 \frac{({\mathcal{D}^{\chi}_{s}} )^2}
 { \varepsilon_s ^5}  
 \left ( \boldsymbol{w}^{\chi}_{s} + \mathbf{W}^{\chi}_{s} \right )   
 \left [ \left ( \varepsilon_s \right )^2 
 - \lambda_s^{\chi} 
 \right ]^2 \, f^\prime_0 (\xi^{\chi}_{s})  \, 
\frac{ \left  (\mathbf k \cross \mathbf{B} \right )  \cdot \mathbf{E}
} {k} \,  ,
\end{gather}
leading to the conductivity components of
\begin{align}
\label{eqsigLF}
(\sigma_s^{\chi, \rm LF})_{i j} = - \,\epsilon_{j q r} \, e^3 \, 
\tau ^2 \, s^3 \, v_0^3 
\int \frac{ d^3 \mathbf{k}} {(2 \, \pi)^3} 
\, \frac{({\mathcal{D}^{\chi}_{s}} )^2}
{  \varepsilon_s ^5} \,
 \left[ (w^{\chi}_{s})_i + (W^{\chi}_{s})_i \right ]   \,
  \left ( \varepsilon_s ^2 - \lambda_s^{\chi}\right )^2  
\, \frac{B_r \,
  k_q \, f^\prime_0 (\xi^{\chi}_{s}) } {k} \,. 
\end{align}
Clearly, the Lorentz-force contribution starts with a linear-in-$B$ term. 
One can also check that $\left ( \sigma_s^{\chi, \rm LF} \right )_{i i} = 0$.

\section{Magnetothermoelectric conductivity} 
\label{appalpha}

The in-plane thermoelectric current density
\begin{align}
	\mathbf{ \bar J}^{\chi}_{ s} &= 
	e \, \tau  \int \frac{d^3 \mathbf{k}} {(2 \, \pi)^3}
	  \, \mathcal{D}^{\chi}_{ s} \left  (\xi^{\chi}_{ s} -  \mu \right )  
	\left [ \boldsymbol{w}^{\chi}_{s} \,+ \, \boldsymbol{W}^{\chi}_{s} \right ]
\left[    \left (  \boldsymbol{w}^{\chi}_{s} +  \boldsymbol{W}^{\chi}_{s} \right )  
\cdot \frac{(-\nabla_{\mathbf{r}} T )}{T} \right ] 
	  f^\prime_0 (\xi^{\chi}_{ s})
\end{align}
gives us the LTEC and TTEC as
\begin{align}
\left(\bar \alpha^{\chi}_{s} \right)_{i j} 
= \,e \, \tau  
\int \frac{ d^3 \mathbf k}{(2\, \pi)^3 } \, \mathcal{D}^{\chi}_{s} 
\left[  (w^{\chi}_{s})_i \, + (W^{\chi}_{s})_i \right ]
\left [ (w^{\chi}_{s})_j \, + (W^{\chi}_{s})_j \right] \,
\frac{(\xi^{\chi}_{s} -  \mu )}{T} \, f^\prime_0 (\xi^{\chi}_{s})  \,.
\end{align}
We divide it up as
\begin{align}
{\bar \alpha}^{\chi}_{s}  =  
\alpha^{\chi, \text{Drude}}_{s} + \alpha^{\chi, \text{BC}}_{s} 
+ \alpha^{\chi, m}_{s} \,,
\end{align} 
analogous to the case of ${\bar \sigma}^{\chi}_{s}$.

\subsubsection{Drude part}

Explicitly, the Drude contribution is expressed as 
\begin{align}
(\alpha^{\chi, \text{Drude}}_{s})_{i j} =  e \, \tau 
\int \frac{ d^3 \mathbf k}{(2\, \pi)^3 } \, \mathcal{L}_{i j}^{(0)} \, 
 \frac{(\varepsilon_{s} -  \mu )}{T} \,  f^\prime_0  (\varepsilon_{s} ) 
 \,,
\quad \mathcal{L}_{i j}^{(0)} =  (v^{\chi}_{ s })_{i} \, (v^{\chi}_{ s })_{j}\,.
\end{align}
The isotropy of the RSW bands, in the vicinity of a node, ensures that the off-diagonal terms vanish, i.e., $\left ( \alpha^{\chi, \text{Drude}}_{s} \right )_{i j} \propto \delta_{ij} $. This leaves only the longitudinal components of the tensor, which are given by
\begin{align}
(\alpha^{\chi, \text{Drude}}_{s})_{i i} = e \, \tau 
\int \frac{ d^3 \mathbf k}{(2\, \pi)^3 } \, [ (v^{\chi}_{ s })_{i} ]^2 \,  
\frac{(\varepsilon_{s} -  \mu )}{T} \,  f^\prime_0  (\varepsilon_{s} )\, . 
\end{align}

\subsubsection{BC-only part (no OMM)}

The BC-only part is given by 
\begin{align}
(\alpha^{\chi, \text{BC}}_{s})_{i j} &=  e \, \tau 
\int \frac{ d^3 \mathbf k}{(2\, \pi)^3 } \,  \mathcal{M}_{i j}   \, \,  \frac{(\varepsilon_{s} -  \mu )}{T} f^\prime_0  (\varepsilon_{s} ) \, 
\end{align}
where $\mathcal{M}_{i j} $ has been defined in Eq.~\eqref{eqmij}. 

\subsubsection{Part with the integrand proportional to nonzero powers of OMM}

The OMM-induced part is given by
\begin{align}
	(\alpha^{\chi, m }_{s})_{i j} & =  \frac{e \, \tau }{ T} 
	\int \frac{d^3 \mathbf{k}}{(2 \, \pi)^3}  \, 
\Bigg [ 
	\left(    \mathcal{P}_{i j} + \mathcal{Q}_{i j}  
	+ \mathcal{R}_{i j} \right) \, \left \lbrace f^\prime_0  (\varepsilon_{s}) + (\varepsilon_{s} -  \mu) \, f^{\prime \prime}_0  (\varepsilon_{s})  \right \rbrace
+ \mathcal{S}_{i j}  \, (\varepsilon_{s} -  \mu)   \, f^\prime_0  (\varepsilon_{s} ) 
\nn & \hspace{ 3 cm}
+ \mathcal{T}_{i j}   \left \lbrace  f^{\prime \prime}_0 (\varepsilon_{s} ) 
+ \frac{1}{2} \, (\varepsilon_{s} -  \mu) f^{\prime \prime \prime}_0 (\varepsilon_{s} ) \right \rbrace
\Bigg ] \,,
\end{align}
where $ \mathcal{P}_{i j}$, $ \mathcal{Q}_{i j}$, $ \mathcal{R}_{i j}$,  $ \mathcal{S}_{i j}$, and $ \mathcal{T}_{i j}$ have been defined in Eq.~\eqref{eqpqrs}.

\section{Magnetothermal coefficient} 
\label{appell}

The in-plane thermal current density is
\begin{align}
\label{eqellcur}
{\mathbf{\bar J}}^{\rm th,\chi}_s & =
	- \,\tau  \int \frac{d^3 \mathbf{k} }  {(2 \, \pi)^3} \, \mathcal{D}^{\chi}_{s}  
	\left (\xi^{\chi}_{s} - \mu \right )^2  
	\left ( \boldsymbol{w}^{\chi}_{ s} +  \boldsymbol{W}^{\chi}_{ s} \right ) 
\left[  \left ( \boldsymbol{w}^{\chi}_{ s} +  \boldsymbol{W}^{\chi}_{ s}\right ) 
\cdot \frac{(- \nabla_{\mathbf{r}} T )}{T} \right ] 
f^\prime_0 (\xi^{\chi}_{\tilde s})\, ,
\end{align}  
which leads to the magnetothermal coefficients expressed as
\begin{align}
\label{eqell}
\left ( {\bar \ell}^{\chi}_{s} \right )_{ij}
= - \,  \tau   \int \frac{d^3 \mathbf{k} }  {(2 \, \pi)^3} \,
\mathcal{D}^{\chi}_{s} \, \frac{  (\xi^{\chi}_{s} - \mu )^2}{T} 
\left[  (w^{\chi}_{ s})_i +  (W^{\chi}_{ s})_i \right ] 
\left[ (w^{\chi}_{ s})_j +  (W^{\chi}_{ s})_j \right ]
f^\prime_0 (\xi^{\chi}_{ s})\,.
\end{align}  

We divide it up as
\begin{align}
{\bar \ell}^{\chi}_{s}  =  
\ell^{\chi, \text{Drude}}_{s} + \ell^{\chi, \text{BC}}_{s} 
+ \ell^{\chi, m}_{s} \,,
\end{align} 
analogous to the cases of ${\bar \sigma}^{\chi}_{s}$ and ${\bar \alpha}^{\chi}_{s}$.

\subsubsection{Drude part}

The Drude contribution is given by 
\begin{align}
(\ell^{\chi, \text{Drude}}_{s})_{i j} = 
- \, \tau  \int \frac{d^3 \mathbf{k} }  {(2 \, \pi)^3} \, \mathcal{L}_{i j}^{(0)} \, 
 \frac{ \left(\varepsilon_{s} -  \mu \right)^2}{T} \,  f^\prime_0  (\varepsilon_{s} ) \,,
\quad \mathcal{L}_{i j}^{(0)} =  
(v^{\chi}_{ s })_{i} \, (v^{\chi}_{ s })_{j}\,.
\end{align}
The isotropy of the RSW bands, in the vicinity of a node, ensures that the off-diagonal terms vanish, i.e., $\left ( \ell^{\chi, \text{Drude}}_{s} \right )_{i j} \propto \delta_{ij} $. This leaves only the longitudinal components of the tensor, which are given by
\begin{align}
(\ell^{\chi, \text{Drude}}_{s})_{i i} = -\, \frac{ \tau }{(2 \, \pi)^3} 
\int d^3 \mathbf{k} \, \left[ (v^{\chi}_{ s })_{i} \right]^2 \,  
\frac{(\varepsilon_{s} -  \mu )^2}{T} \,   f^\prime_0  (\varepsilon_{s} )\, . 
\end{align} 

\subsubsection{BC-only part (no OMM)}

The BC-only part is given by 
\begin{align}
(\ell^{\chi, \text{BC}}_{s})_{i j} &= 
- \,\tau  \int \frac{d^3 \mathbf{k} } 
{(2 \, \pi)^3} \,  \mathcal{M}_{i j}   \, 
  \frac{(\varepsilon_{s} -  \mu )^2}{T} 
  \,f^\prime_0  (\varepsilon_{s} )\, ,  
\end{align}
where $\mathcal{M}_{i j} $ is defined in Eq.~\eqref{eqmij}.

\subsubsection{Part with the integrand proportional to nonzero powers of OMM}

The OMM-induced part is captured by 
\begin{align}
(\ell^{\chi,m }_{s})_{i j} &= 
- \, \frac{ \tau }{ T} 
		\int \frac{d^3 \mathbf{k}}{(2 \, \pi)^3} \,\Big[  \mathcal{T}_{i j}
\left \lbrace f^\prime_0 
\left(\varepsilon_{s} \right) 
+ 2 \left(\varepsilon_{s} -  \mu \right) \, f^{\prime \prime}_0 
\left(\varepsilon_{s} \right) 
+  \frac{1}{2} 
 \left(\varepsilon_{s} -  \mu \right)^2 
f^{\prime \prime \prime}_0 \left(\varepsilon_{s} \right)   
\right \rbrace
+  \mathcal{S}_{i j} 
\left(\varepsilon_{s} -  \mu \right)^2\, f^\prime_0 (\varepsilon_{s} ) \nonumber \\
&  \hspace{ 3 cm}
+ \left(\mathcal{P}_{i j} + \mathcal{Q}_{i j}  +  \mathcal{R}_{i j} \right)  
 \left(\varepsilon_{s} -  \mu \right)\, 
\left \lbrace 2 \,  f^\prime_0 (\varepsilon_{s} ) 
	+   \left (\varepsilon_{s} -  \mu \right)\,   f^{\prime \prime}_0 (\varepsilon_{s} ) 
	\right \rbrace \Big]\,,
\end{align}
where $\mathcal{P}_{i j}$, $\mathcal{Q}_{i j}$, $\mathcal{R}_{i j}$, $\mathcal{S}_{i j}$, and $\mathcal{T}_{i j}$ have been defined in Eq.~\eqref{eqpqrs}.


\bibliography{ref_rsw}
\end{document}